\documentclass[12pt]{article}
\usepackage[T1]{fontenc}
\usepackage{cite}
\usepackage{mathrsfs}
\usepackage{amsmath}
\usepackage{amssymb}
\usepackage[dvipdfm]{graphicx}
\usepackage{tabularx}
\usepackage{bm}

\allowdisplaybreaks[4]

\makeatletter
\@addtoreset{equation}{section}

\makeatletter
\renewcommand\section{\@startsection {section}{1}{\z@}%
                                   {-3.5ex \@plus -1ex \@minus -.2ex}
                                   {2.3ex \@plus.2ex}%
                                   {\normalfont\large\bfseries}}
\renewcommand\subsection{\@startsection{subsection}{2}{\z@}%
                                     {-3.25ex\@plus -1ex \@minus -.2ex}%
                                     {1.5ex \@plus .2ex}%
                                    {\normalfont\bfseries}}

\begin{document}

\begin{titlepage}
  \thispagestyle{empty}
  
  \begin{flushright} 
    KUNS-2170\\
    
  \end{flushright} 
  
  \vspace{2cm}
  
  \begin{center}
    \font\titlerm=cmr10 scaled\magstep4
    \font\titlei=cmmi10 scaled\magstep4
    \font\titleis=cmmi7 scaled\magstep4
     \centerline{\titlerm
       Instabilities of Kerr-AdS{\LARGE $_{5}\times S^5$} Spacetime}
    
    \vspace{2.2cm}
    \noindent{{
        \large{Keiju Murata}
      }}\\
    \vspace{0.8cm}
    
   {\it Department of Physics, Kyoto University, Kyoto 606-8501, Japan}
   \\
   e-mail:murata@tap.scphys.kyoto-u.ac.jp
    
   \vspace{1cm}
   {\large \today}
  \end{center}

  \vskip 3em

  \begin{abstract}
We study gravitational perturbations of the Kerr-AdS$_5\times S^5$ spacetime
with equal angular momenta. 
In this spacetime, we found the two kinds of classical instabilities, 
superradiant and Gregory-Laflamme instabilities.
The superradiant instability is caused by the wave
amplification via superradiance, and by wave reflection 
due to the potential barrier of the AdS spacetime. 
The Gregory-Laflamme instability appears in Kaluza-Klein modes of
the internal space $S^5$ and breaks the symmetry $SO(6)$. 
Taken into account these instabilities, 
the phase structure of Kerr-AdS$_5\times S^5$ spacetime is revealed.
The implication for the AdS/CFT correspondence is also discussed. 
  \end{abstract}

\end{titlepage}

 \noindent\rule\textwidth{.1pt}
 \vskip 2em \@plus 3ex \@minus 3ex
 \tableofcontents
 \vskip 2em \@plus 3ex \@minus 3ex
 \noindent\rule\textwidth{.1pt}
 \vskip 2em \@plus 3ex \@minus 3ex

\section{Introduction and Summary}
Recently, AdS black holes in $S^5$ compactified type IIB supergravity have
attracted much interest because
they describe strongly coupled
$\mathcal{N}=4$ thermal super Yang-Mills theory via AdS/CFT 
correspondence~\cite{Maldacena:1997re,Gubser:1998bc,Witten:1998qj,Aharony:1999ti}. 
Especially, phase transitions of dual gauge theory are identified with 
instabilities of AdS black holes
and understanding of stability of
AdS black holes is important to reveal the strongly coupled gauge
theory.
The stability of Schwarzschild-AdS black holes
has been shown
in~\cite{Kodama:2003jz,Ishibashi:2003ap,Kodama:2003kk,Konoplya:2008rq,Konoplya:2007jv}. 
However, in the case of Kerr-AdS black holes, we can expect an instability
called superradiant instability. 
The perturbation of Kerr-AdS black holes can be
amplified by superradiance at the horizon. 
On the other hand, at the infinity, the
amplified perturbation will be reflected by the potential barrier of the AdS
spacetime. This will be amplified at the horizon again. By repetition of this
mechanism, the initial perturbation can grow exponentially and 
Kerr-AdS black holes become unstable. 
The superradiant instability is physically reasonable, but, practically, it is
difficult to find the instability by gravitational perturbation because
of the difficulty of separation of perturbation equations.
Nevertheless, in some special cases, 
there are several works on the stability of Kerr-AdS
black holes. In the case of 4-dimensional Kerr-AdS spacetime, the superradiant 
instability has been found~\cite{Cardoso:2004hs,Cardoso:2006wa}.
In $D=7,9,11,\cdots$, 
the same instability of Kerr-AdS black holes with equal angular momenta 
has been shown to exist~\cite{Kunduri:2006qa}. 
In the case of $(D\geq 7)$-dimensional Kerr-AdS black hole with one rotating axis, 
it has been shown that the superradiant instability  appears
 in the tensor type perturbation~\cite{Kodama:2007sf,Kodama:2008rq}. 
However, there is no work for the stability analysis of 
five-dimensional Kerr-AdS black holes (except for a massless Kerr-AdS
black holes~\cite{Carter:2005uw} or a scalar field
perturbation~\cite{Aliev:2008yk}).
To get relevant results for the AdS$_5$/CFT$_4$ correspondence, 
we need to study the instability of five-dimensional Kerr-AdS black
holes. 
It is difficult to study the stability of the general Kerr-AdS$_5$
spacetime. However, for equal angular momenta case, 
spacetime symmetry of Kerr-AdS$_5$ black hole is enhanced and the
separation of gravitational perturbation equations can be
possible~\cite{Murata:2007gv,Murata:2008yx,Kimura:2007cr,Ishihara:2008re}. 
One of our purposes is to find the superradiant instability of
five-dimensional Kerr-AdS black holes with equal angular momenta.

The superradiant instability is caused by property of rotating AdS
black holes and information of the internal space $S^5$ is not so important
for superradiant instability.
However, if the internal space $S^5$ is taken into account, 
we can find another type of instability, called Gregory-Laflamme
instability. Originally, the Gregory-Laflamme instability has been
found in the black brane solution~\cite{Gregory:1993vy,Gregory:1994bj,Harmark:2007md}, 
but, in the Schwarzschild-AdS$_5\times S^5$ spacetime, 
the situation can be similar to the black brane system. If the horizon
radius is much smaller than radius of $S^5$, the internal space may be
considered as $\bm{R}^5$. Then, we can consider Sch-AdS$_5\times S^5$
spacetime as a black brane and the Gregory-Laflamme instability may
appear in Kaluza-Klein modes. 
The Gregory-Laflamme instability of Schwarzschild-AdS$_5\times S^5$ spacetime
has been already found in~\cite{Hubeny:2002xn}.
In this paper, extending their work, we study the Gregory-Laflamme instability of
Kerr-AdS$_5\times S^5$ spacetime. 

We will take into account Gregory-Laflamme and superradiant instabilities
and reveal the phase structure of Kerr-AdS$_5\times S^5$ spacetime.
There are two kinds of instabilities in the Kerr-AdS$_5\times S^5$ spacetime.
Thus, we can expect that this spacetime has rich phase structure and it
will be useful to find the evidence of the AdS/CFT correspondence. 



\begin{figure}
  \begin{center}
      \includegraphics[height=6cm,clip]{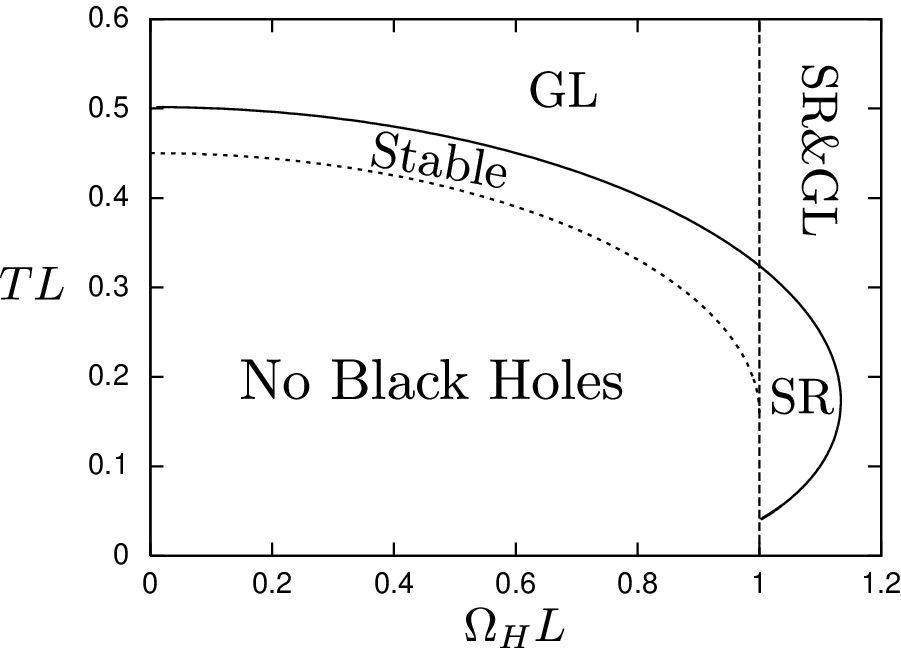}
    \caption{\label{fig:phase}This is the Phase diagram of a
   small Kerr-AdS$_5\times S^5$ black hole. 
The values $\Omega_H$ and $T$ are angular velocity and temperature of Kerr-AdS$_5$
   black holes. These are normalized by the curvature scale of the AdS
   spacetime, $L$. 
In the ``Stable'' region, Kerr-AdS
   black holes are stable. In the ``SR'' and ``GL'' region, black holes are
   unstable against superradiant and Gregory-Laflamme instability,
   respectively. In ``SR{\&}GL'' region, black holes are unstable
   against both of them. In ``No Black Holes'' region, there is no black
   hole solution.
}
  \end{center}
\end{figure}

The organization and summary of this paper is as follows. 
In section~\ref{sec:KerrAdS}, we introduce Kerr-AdS$_5\times S^5$
spacetimes with equal angular momenta. Especially, the spacetime
symmetry is studied. We shall see that the symmetry is 
$R_t\times SU(2)\times U(1)\times SO(6)$ in the case of equal angular
momenta.
In section~\ref{sec:SR}, 
we study the gravitational perturbation of Kerr-AdS$_5$ spacetime neglecting the
Kaluza-Klein modes of the internal space $S^5$. 
We can get the master equations which are relevant for the superradiant
instability. These equations are solved numerically and 
we find the onset of superradiant instabilities given by 
$\Omega_HL=1$, where $\Omega_H$ is angular velocity of horizon and $L$
is curvature scale of AdS spacetime. 
In section~\ref{sec:GL},
we study Gregory-Laflamme instability of Kerr-AdS$_5\times S^5$
spacetime. We consider the gravitational perturbation 
including Kaluza-Klein modes in order to see the Gregory-Laflamme instability
and get the ordinary differential equations in which three variables are
coupled. These equations are solved numerically and we find the 
Gregory-Laflamme instability. 
In section~\ref{sec:Phase},
we reveal the phase structure of Kerr-AdS$_5\times S^5$ spacetime 
taking into account superradiant and Gregory-Laflamme instabilities.
The result is in Figure.\ref{fig:phase},
$\Omega_H$ and $T$ are angular velocity and temperature of Kerr-AdS$_5$
black holes. In Figure.\ref{fig:phase}, $\Omega_H$ and $T$ are
normalized by the curvature scale of 
the AdS spacetime, $L$. 
The solid and dashed lines are onset of the Gregory-Laflamme and
superradiant instabilities, respectively. These lines cross each other
and we can see five phases in this diagram. 
In the ``Stable'' region, Kerr-AdS black holes are stable. 
In the ``SR'' and ``GL'' region, black holes are
unstable against superradiant and Gregory-Laflamme instabilities,
respectively. In ``SR{\&}GL'' region, black holes are unstable
against both of them. In ``No Black Holes'' region, there is no black
hole solution.
The final section is devoted to the conclusion.

\section{Kerr-AdS$_5$ black hole in Type IIB Supergravity}\label{sec:KerrAdS}
 
\subsection{Kerr-AdS$_5\times S^5$ spacetime with equal angular momenta}
In this section, we introduce Kerr-AdS$_5$ spacetime
as a solution of type IIB supergravity. 
The equations of motion of type IIB supergravity are
given by
\begin{gather}
R_{MN}=\frac{1}{48}F_{MP_2P_3P_4P_5}F_N{}^{P_2P_3P_4P_5}
-\frac{1}{480}g_{MN}F_{P_1P_2P_3P_4P_5}F^{P_1P_2P_3P_4P_5}\ ,\label{eq:BGEOM1}\\
\nabla_{P_1}F^{P_1P_2P_3P_4P_5}=0\ ,
\label{eq:BGEOM2}
\end{gather}
where $M,N,\cdots=0,1,\cdots,9$. The form $F_{M_1M_2M_3M_4M_5}$ is RR 5-form
satisfying $d\bm{F}=0$ and $\ast \bm{F} = \bm{F}$. 
We concentrate on the metric and RR 5-form field in type IIB
supergravity, while
other components, such as dilaton, NSNS 3-form,
RR 1-form and 3-form, have been set to be zero.
We will consider Kerr-AdS$_5\times S^5$ spacetime
which is a solution of~(\ref{eq:BGEOM1}) and (\ref{eq:BGEOM2}).
The Kerr-AdS$_5$ spacetime can have two independent angular momenta
generally, but, for simplicity, we will consider the case of equal two
angular momenta. Then, the spacetime symmetry is enhanced and stability
analysis will be possible.
The metric of Kerr-AdS$_5\times S^5$ spacetime with equal angular momenta is given 
by\footnote{To obtain this metric from Kerr-AdS$_5$ spacetime 
given in~\cite{Hawking:1998kw,Gibbons:2004js,Gibbons:2004uw,Gibbons:2004ai},
we need some coordinate transformation and redefinition of a parameter.
These are summarized in appendix~\ref{app:KerrAdS_indep}.
}
\begin{multline}
  ds^2= - \left(1+\frac{ r^2}{L^2}\right) dt^2
+\frac{d r^2}{G( r)}+
\frac{ r^2}{4} \{(\sigma^1)^2+(\sigma^2)^2+(\sigma^3)^2\}\\
+\frac{2\mu}{ r^2} \left(dt+\frac{a}{2}\sigma^3\right)^2
+L^2 d\Omega_5^2\ ,
\label{eq:KerrAdS}
\end{multline}
where $G(r)$ is defined by
\begin{equation}
 G( r) = 1+\frac{ r^2}{L^2} 
- \frac{2\mu( 1- a^2/L^2)}{ r^2}+\frac{2\mu a^2}{ r^4}\ .
\end{equation}
Then, RR 5-form is
\begin{equation}
 \bm{F}=
2^{3/2}L^{-1} (\bm{\epsilon}_{\text{AdS}_5}+\bm{\epsilon}_{S^5})
\label{eq:F}
\end{equation}
where $\bm{\epsilon}_{S^5}$ is volume form of $L^2d\Omega_5^2$ and
$\bm{\epsilon}_{\text{AdS}_5}$ is volume form of AdS$_5$ part
of~(\ref{eq:KerrAdS}). Because of the relation,
$\ast\bm{\epsilon}_{S^5}=\bm{\epsilon}_{\text{AdS}_5}$, the form $\bm{F}$ satisfies
the self dual condition.
%
In (\ref{eq:KerrAdS}), we have defined the invariant forms
$\sigma^a\,(a=1,2,3)$ of $SU(2)$ as
\begin{equation}
 \begin{split}
  \sigma^1 &= -\sin\psi d\theta + \cos\psi\sin\theta d\phi\ ,\\
  \sigma^2 &= \cos\psi d\theta + \sin\psi\sin\theta d\phi\ ,\\
  \sigma^3 &= d\psi + \cos\theta d\phi  \ ,
 \end{split}
\end{equation}
where  $0\leq \theta < \pi $, $0\leq \phi <2\pi$, $0\leq \psi <4\pi$.
It is easy to check the relation 
$d\sigma^a = 1/2 \epsilon^{abc} \sigma^b \wedge \sigma^c$. 
The dual vectors of $\sigma^a$ are given by
\begin{equation}
\begin{split}
 e_1 &= -\sin\psi \partial_\theta +
  \frac{\cos\psi}{\sin\theta}\partial_\phi - \cot\theta\cos\psi
 \partial_\psi\ ,\\
 e_2 &= \cos\psi \partial_\theta +
  \frac{\sin\psi}{\sin\theta}\partial_\phi - \cot\theta\sin\psi
 \partial_\psi\ ,\\
 e_3 &= \partial_\psi\ ,
\end{split}
\end{equation}
and, by the definition, they satisfy 
$\sigma^a_i e^i_b =\delta^a_b$.  

The horizon radius $r=r_+$ can be determined by $G(r_+)=0$. The
angular velocity of Kerr-AdS back hole is given by
\begin{equation}
 \Omega_H = \frac{2\mu a}{ r_+^4 + 2\mu a^2}\ .
\label{eq:degen_angular_vel}
\end{equation}
For existence of horizon, the angular velocity has the upper bound,
\begin{equation}
 \Omega_H \leq
  \left(\frac{1}{2r_+^2}+\frac{1}{L^2}\right)^{1/2}
 \equiv \Omega_H^\text{max}\ .
\label{eq:Omegamax}
\end{equation}
In term of $ r_+$ and $\Omega_H$, the two parameters $(a,\mu)$ in the
metric~(\ref{eq:KerrAdS}) can be rewritten as
\begin{equation}
a = \frac{ r_+^2 \Omega_H}{1+ r_+^2/L^2}\ ,\quad
\mu=
 \frac{1}{2}\frac{ r_+^2(1+ r_+^2/L^2)^2}{1-(\Omega_H^2L^2-1) r_+^2/L^2}\ .  
\label{eq:mu_a_to_r+_Omega}
\end{equation}
We will use parameters $(r_+,\Omega_H)$ mainly.

\subsection{Spacetime Symmetry}
Now, we study the symmetry of~(\ref{eq:KerrAdS}). 
It will be important for separation of variables of the gravitational
perturbation equations in section \ref{sec:SR} and \ref{sec:GL}.
Apparently, 
the metric~(\ref{eq:KerrAdS}) has the time translation symmetry $R_t$
and $SO(6)$ symmetry comes from $S^5$ part of~(\ref{eq:KerrAdS}).
Additionally, the spacetime has the $SU(2)$ symmetry
characterized by the Killing vectors $\xi_{\alpha} \ , (\alpha=x,y,z)$:
\begin{equation}
\begin{split}
 \xi_x &= \cos\phi\partial_\theta +
 \frac{\sin\phi}{\sin\theta}\partial_\psi -
 \cot\theta\sin\phi\partial_\phi\ ,\\
 \xi_y &= -\sin\phi\partial_\theta +
 \frac{\cos\phi}{\sin\theta}\partial_\psi -
 \cot\theta\cos\phi\partial_\phi\ ,\\
 \xi_z &= \partial_\phi\ .
\end{split}
\end{equation}
The symmetry can be explicitly shown by using
 the relation $\mathcal{L}_{\xi_{\alpha}}\sigma^a=0$,
 where $\mathcal{L}_{\xi_{\alpha}}$ is a Lie derivative
 along the curve generated by the vector field $\xi_{\alpha}$.

From the metric (\ref{eq:KerrAdS}), we can also read off 
the additional $U(1)$ symmetry, which 
keeps the part of the metric, $(\sigma^1)^2+(\sigma^2)^2$ and is
generated by $e_3$. 
The $U(1)$ generator $e_3$ satisfies 
$\mathcal{L}_{e_3}\sigma^1=-\sigma^2$ and
$\mathcal{L}_{e_3}\sigma^2=\sigma^1$ and, 
thus, $\mathcal{L}_{e_3}[(\sigma^1)^2+(\sigma^2)^2]=0$.
Therefore, the symmetry of
Kerr-AdS$_5\times S^5$ spacetime with equal angular momenta 
becomes $R_t\times SU(2)\times U(1)\times SO(6)$.  

For later calculations, it is convenient to 
define the new invariant forms
\begin{equation}
 \sigma^{\pm} = \frac{1}{2}(\sigma^1 \mp i \sigma^2)\ . 
\end{equation}
%
%
Then, the dual vectors for $\sigma^\pm$ are
\begin{equation}
 \bm{e}_{\pm} = \bm{e}_1 \pm i \bm{e}_2\ .
\end{equation}
By making use of these forms, the metric~(\ref{eq:KerrAdS}) can be rewritten as
\begin{equation} 
ds^2= - \left(1+\frac{ r^2}{L^2}\right) dt^2
+\frac{d r^2}{G( r)}+
\frac{ r^2}{4} \{4\sigma^+\sigma^- +(\sigma^3)^2\}
+\frac{2\mu}{ r^2} \left(dt+\frac{a}{2}\sigma^3\right)^2
+L^2 d\Omega_5^2\ .
\label{eq:KerrAdS2}
\end{equation}
We will use this expression in the following sections.

\section{Superradiant Instability of Kerr-AdS black holes}
\label{sec:SR}

In the following sections, we will study the stability of 
Kerr-AdS$_5\times S^5$ spacetime with equal angular momenta~(\ref{eq:KerrAdS}).
In this spacetime, we can expect two kinds of instabilities.
One of them is the superradiant instability which is caused by the wave
amplification via superradiance, and by wave reflection 
due to the potential barrier of the AdS spacetime. 
This instability
should be seen, even if Kaluza-Klein modes are neglected.
The other instability is the Gregory-Laflamme instability which is
instability of the internal space $S^5$, that is to say, the
Gregory-Laflamme instability is instability of Kaluza-Klein modes.
First, we shall see the superradiant instability of
Kerr-AdS$_5$ spacetime in this section. 

\subsection{Perturbation equations and separability}
\label{sec:formalism}

To see the superradiant instability, we can neglect Kaluza-Klein modes of
$S^5$. In addition, we will consider only  metric
fluctuation on the AdS$_5$ part of the spacetime, that is, 
\begin{equation}
\begin{split}
 &g'_{MN}dx^Mdx^N=g_{MN}dx^Mdx^N+h_{\mu\nu}(x^\mu)dx^\mu dx^\nu\ ,\\
 &\bm{F}'=2^{3/2}L^{-1} (\bm{\epsilon}'_{\text{AdS}_5}+\bm{\epsilon}_{S^5})
\end{split}
\label{eq:fluc0}
\end{equation}
where $\mu,\nu\cdots$ are indexes on AdS$_5$ and 
$\bm{\epsilon}'_{\text{AdS}_5}$ is volume form of 
$g'_{\mu\nu}=g_{\mu\nu}+h_{\mu\nu}$.
For the perturbations, the equation~(\ref{eq:BGEOM2}) is trivially
satisfied and equation~(\ref{eq:BGEOM1}) becomes 
\begin{equation}
 \delta G_{\mu\nu}-\frac{6}{L^2}h_{\mu\nu}=0\ .
\label{eq:tensor_EOM}
\end{equation}
where $\delta G_{\mu\nu}$ is 
perturbation of the Einstein tensor of five-dimensional metric $g_{\mu\nu}$,
which is defined by
\begin{multline}
 \delta G_{\mu\nu}=\frac{1}{2}[\nabla^\rho\nabla_\mu h_{\nu \rho}
 + \nabla^\rho\nabla_\nu h_{\mu\rho} -
  \nabla^2 h_{\mu\nu} - \nabla_\mu\nabla_\nu h\\
-g_{\mu\nu}(\nabla^\rho\nabla^\sigma h_{\rho\sigma} - \nabla^2 h -
 R^{\rho\sigma}h_{\rho\sigma}) - R h_{\mu\nu}]\ ,
\label{eq:delG}
\end{multline}
where $\nabla_\mu$ denotes the covariant derivative with respect to $g_{\mu\nu}$
and $h= g^{\mu\nu} h_{\mu\nu}$. Tensors $R_{\rho\sigma}$ and $R$ are Ricci
tensor and Ricci scalar of $g_{\mu\nu}$. 
We take AdS$_5$ part of~(\ref{eq:KerrAdS2}) as a background metric $g_{\mu\nu}$.
The equation~(\ref{eq:tensor_EOM}) is nothing but the perturbation of five-dimensional Einstein
equations with the negative cosmological constant. 


The perturbation equation~(\ref{eq:tensor_EOM}) is a partial differential
equation of $h_{\mu\nu}(t,r,\theta,\phi,\psi)$.
However, in previous 
works~\cite{Hu:1974hh,Murata:2007gv,Kimura:2007cr,Murata:2008yx},
it was shown that the perturbation equations can be reduced to ordinary
differential equations by focusing on the symmetry of the background spacetime, 
$R_t\times SU(2)\times U(1)$.
Here, we will briefly review these works.

Let us define the two kinds of angular momentum operators
\begin{equation}
  L_\alpha = i \xi_\alpha \ , \quad
  W_a  = i  {e}_a \ .
\end{equation}
where $\alpha,\beta,\cdots = x,y,z$ and $a,b,\cdots = 1,2,3$. 
They satisfy commutation relations
\begin{equation}
 [L_\alpha, L_\beta] = i \epsilon_{\alpha\beta\gamma} L_\gamma \  , \quad
 [W_a, W_b] = -i\epsilon_{abc} W_c
 \ , \quad [L_\alpha, W_a]=0 \ ,
\end{equation}
where $\epsilon_{\alpha\beta\gamma}$ and $\epsilon_{abc}$ are
antisymmetric tensors which satisfy $\epsilon_{123}=\epsilon_{xyz}=1$. 
The Casimir operators constructed by $L_\alpha$ and $W_a$ are identical and
we define $L^2 \equiv L_\alpha^2 = W_a^2$.
The symmetry group, $SU(2)\times U(1)$ is generated by $L_\alpha$ and
$W_3$.
Here, we should notice the fact
\begin{eqnarray}
\mathcal{L}_{W_3} \sigma^{\pm}
=  \pm  \sigma^{\pm} \ ,\quad 
\mathcal{L}_{W_3} \sigma^3= 0\ .
 \label{rule}
\end{eqnarray}
It means that  $\sigma^{\pm}$ and $\sigma^3$ have $U(1)$ charges $\pm 1$ and $0$. 
Since operators $L^2$, $L_z$ and $W_3$ commute each other,
these are simultaneously diagonalizable. 
The eigenfunctions
are called Wigner functions $D^J_{KM}(\theta,\phi,\psi)$
defined by 
\begin{equation}
 L^2 D^J_{KM} = J(J+1)D^J_{KM}\ ,\quad
 L_z D^J_{KM} = M D^J_{KM}\ ,\quad
 W_3 D^J_{KM} = K D^J_{KM}\ ,
\label{eq:WigDef}
\end{equation}
where indexes $J,K$ and $M$ are defined for 
$J=0,1/2,1,\cdots$ and $K,M=-J,-J+1,\cdots J$.
%
%
%
The following relations are useful for later calculations
\begin{equation}
 W_+ D^J_{KM} = i \epsilon_K D^J_{K-1,M}\ ,\quad W_- D^J_{KM} = -i\epsilon_{K+1}D^J_{K+1,M}\
  , \quad W_3 D^J_{KM} = KD^J_{KM}\ ,
\end{equation}
where we have defined $W_\pm = W_1\pm i W_2$ and $\epsilon_K=\sqrt{(J+K)(J-K+1)}$. 
From this relation, we get the differential rule of the Wigner
function as
\begin{equation}
 \partial_+ D^J_{KM} = \epsilon_K D^J_{K-1,M}\ ,\quad
 \partial_- D^J_{KM} = -\epsilon_{K+1}D^J_{K+1,M}\ ,\quad
 \partial_3 D^J_{KM} = -iKD^J_{KM}\ ,
\label{eq:delD}
\end{equation}
where we have defined 
$\partial_\pm \equiv e_\pm^i\partial_i$ and $ \partial_3 \equiv e_3^i\partial_i$.

Now, we consider the mode expansion of $h_{\mu\nu}$. 
The metric perturbations can be classified into three parts,
$h_{AB},h_{Ai},h_{ij}$ where $A,B=t,r$ and $i,j=\theta,\phi,\psi$. 
They behave as scalar, vector and tensor for
coordinate transformation of $\theta,\phi,\psi$, respectively. 
The scalar
$h_{AB}$ can be expanded by Winger functions immediately as 
\begin{equation}
 h_{AB} = \sum_{K} h_{AB}^K(x^A) D_K(x^i)\ .
\label{eq:hAB_dec}
\end{equation}
Here, we have omitted the indexes $J,M$ because the differential rule of Wigner
function~(\ref{eq:delD}) cannot shift $J, M$ and therefore the modes with different
eigenvalues $J, M$ are trivially decoupled in the perturbation equations. 

To expand the vector part $h_{Ai}$, we need a device. 
First, we change the basis $\{\partial_i\}$ to
$\{ {e}^a\}$, that is $h_{Ai}=h_{Aa}\sigma^a_i$ where $a=\pm,3$. 
Then, because $h_{Aa}$ is scalar, we can expand it by the Wigner function as
\begin{equation}
\begin{split}
 h_{Ai}(x^\mu)&=h_{A+}(x^\mu)\sigma^+_i + h_{A-}(x^\mu)\sigma^-_i + h_{A3}(x^\mu)\sigma^3_i\\
 & = \sum_K \left[h_{A+}^K(x^A) \sigma^+_i D_{K-1} + h_{A-}^K(x^A)\sigma^-_i D_{K+1}
 + h_{A3}^K(x^A)\sigma^3_iD_K\right]\ .\\
\end{split}
\label{eq:hAi_dec}
\end{equation}
In the expansion of $h_{A+}$, $h_{A-}$ and $h_{A3}$, we have
shifted the index $K$ of Wigner functions, for example, 
$h_{A+}$ has been expanded as $\sum_K h_{A+}^K D_{K-1}$. 
The reason is as follows. 
The invariant forms $\sigma^\pm$ and 
$\sigma^3$ have the $U(1)$ charge $\pm1$ and $0$, respectively 
(see Eq.~(\ref{rule})), while
the Wigner function $D_K$ has the $U(1)$ charge $K$ (see
Eq.~(\ref{eq:WigDef})). 
Therefore, by shifting the index $K$, we can assign the same $U(1)$ charge $K$
to $\sigma^+_i D_{K-1}$, $\sigma^-_i D_{K+1}$ and $\sigma^3_iD_K$ in
Eq.~(\ref{eq:hAi_dec}). 

The expansion of tensor part $h_{ij}$ can be carried out in a
similar way  as
\begin{eqnarray}
 h_{ij}(x^\mu) 
 &=&  \sum_K  \left[ h_{++}^K  \sigma^+_i \sigma^+_j  D_{K-2} 
  +   2 h_{+-}^K\sigma^+_i \sigma^-_j D_{K} 
  +  2 h_{+3}^K\sigma^+_i \sigma^3_j D_{K-1}   \right. \nonumber \\
&& \left.  \qquad   +  h_{--}^K \sigma^-_i \sigma^-_j D_{K+2} 
  + 2 h_{-3}^K\sigma^-_i \sigma^3_j D_{K+1} 
  +   h_{33}^K \sigma^3_i \sigma^3_j D_{K}    \right]     \  .
\label{eq:hij_dec}
\end{eqnarray}
To assign the same $U(1)$ charge $K$  to each term, 
we have shifted the index $K$ of Wigner functions. 

Substituting Eqs.~(\ref{eq:hAB_dec}), (\ref{eq:hAi_dec}),
(\ref{eq:hij_dec}) into the perturbation equations~(\ref{eq:tensor_EOM}),
 we get the equations for each mode labelled by $J$, $M$, $K$. 
Because of $SU(2)\times U(1)$ symmetry, different eigenmodes cannot appear in the
same equation.  

It is interesting that we can find master variables from above information. 
First, we should notice that coefficients of the expansion have different indexes $K$ 
and, therefore, coefficients of components $h_{AB}^K$, $h_{Aa}^K$ and $h_{ab}^K$ are 
restricted as follows:
\begin{equation}
\begin{array}{|c|c|c|c|c|}
\hline
h_{++}      & h_{A+},h_{+3} & h_{AB}, h_{A3} , h_{+-},h_{33}&h_{A-},h_{-3} &h_{--} \\ \hline
|K-2|\leq J &|K-1|\leq J    &|K|\leq J                      &|K+1|\leq J   &|K+2|\leq J \\ \hline
\end{array}
\nonumber
\end{equation}
From this table, we can see that, in $K=\pm(J+2)$ mode, there is only
one variable $h_{\pm\pm}$, respectively. Therefore, the $(J,M,K=\pm(J+2))$
modes always reduce to a single master equation.
We will study the stability of these modes.
In fact, $(J=0,M=0,K=0,\pm1)$ modes also reduce to a single
master equation. 
The stability of $(J=0,M=0,K=0,\pm1)$ modes 
are studied in appendix \ref{app:other} 
and we will see that these modes are irrelevant to see the onset of 
the superradiant instability.

\subsection{Master Equations}\label{sec:MasterEqs}

We will derive the master equation for $(J,M,K=\pm(J+2))$ modes.
Because of the relation
$h_{++}=h_{--}^\ast$, we will consider $(J,M,K=J+2)$ modes only.
Then, we can set $h_{\mu\nu}$ as
\begin{equation}
 h_{\mu\nu}(x^\mu) dx^\mu dx^\nu = h_{++}(r)e^{-i\omega t}D_{J}(x^i)\sigma^+
  \sigma^+ \ ,
\label{eq:h_K=J+2}
\end{equation}
where $D_{J}\equiv D^J_{K=J,M}$. 
This $h_{++}$ field is gauge invariant. 
We substitute Eq.~(\ref{eq:h_K=J+2}) into Eq.~(\ref{eq:tensor_EOM})
and use the differential rule of Wigner functions~(\ref{eq:delD}).
Then, $++$ component of~(\ref{eq:tensor_EOM}) is given by
\begin{equation}
\begin{split}
 \frac{1}{2r^{10}G(r)}\bigg[&
-r^{10}G(r)^2h_{++}''
-r^5G(r)(6\mu r^2 \lambda a^2-10\mu a^2+6\mu r^2-\lambda
 r^6-r^4)h_{++}'\\
&-\big\{
-4\lambda^2 r^{12}
+(4\lambda(3+3J+J^2)+\omega^2)r^{10}
-4(J+1)(J+2)r^8\\
&-2\mu(-4+16\lambda a^2+4J\lambda a^2-12J-4J^2 +4a(J+2) \omega-a^2
 \omega^2)r^6\\
&+8\mu(2\mu+2\mu\lambda^2 a^4+4\mu a^2
\lambda+Ja^2+4a^2) r^4\\
&-48a^2\mu^2(1+\lambda a^2) r^2
+32\mu^2 a^4
\big\}h_{++}
\bigg]e^{-i\omega t}D_J(\theta,\phi,\psi)=0\ .
\end{split}
\end{equation}
This equation can be rewritten as
\begin{equation}
 -\frac{d^2 \Phi}{d r_\ast^2} + V( r)\Phi =
  [\omega-2(J+2)\Omega( r)]^2 \Phi\ ,
\label{eq:master_K=J+2}
\end{equation}
where we have introduced the new variable,
\begin{equation}
 \Phi = \frac{( r^4+2\mu a^2)^{1/4}}{ r^{3/2}}h_{++}\ .
\end{equation}
and the tortoise coordinate,
\begin{equation}
 d r_\ast = \frac{( r^4+2\mu a^2)^{1/2}}{r^2 G(r)}dr\ .
\label{eq:kame}
\end{equation}
The function $\Omega(r)$ and potential $V(r)$ are given by
\begin{equation}
 \Omega( r)=\frac{2\mu a}{ r^4 + 2\mu a^2} ,
\end{equation}
and 
\begin{equation}
\begin{split}
 V( r) =& \frac{G( r)}{4 r^2( r^4+2\mu a^2)^3}
\big[
15 r^{14}/L^2
+ (4J+7)(4J+5) r^{12}
+ 6\mu(3+11a^2/L^2) r^{10}\\
&\quad + 2\mu a^2 (16J^2+32J+5) r^8
-4\mu^2 a^2(10-17a^2/L^2) r^6\\
&\qquad -4\mu^2 a^4(16J+35) r^4
+8\mu^3 a^4(1-a^2/L^2) r^2
-40\mu^3 a^6
\big]\ .
\end{split}
\end{equation}
We can obtain the asymptotic form of $\Omega( r)$ and $V( r)$ as
\begin{equation}
\Omega( r) \rightarrow \Omega_H \quad( r\rightarrow  r_+)\ ,\quad
\Omega( r) \rightarrow 0 \quad( r\rightarrow \infty)\ ,
\end{equation}
and
\begin{equation}
 V( r)\rightarrow 0 \quad( r\rightarrow  r_+)\ ,\quad 
 V( r)\rightarrow 
\frac{15r^2}{4L^4} 
\quad( r\rightarrow \infty)\ ,
\end{equation}
where $\Omega_H$ is the angular velocity of the horizon which is defined
in Eq.~(\ref{eq:degen_angular_vel}). Therefore, the asymptotic form of
the solution of master equation~(\ref{eq:master_K=J+2}) becomes
\begin{equation}
\Phi\rightarrow e^{\pm i\{\omega-2(J+2)\Omega_H\} r_\ast}
 \quad( r\rightarrow  r_+)\ ,\quad
\Phi \rightarrow 
r^{-1/2\pm 2}
\quad( r\rightarrow \infty)\ .
\end{equation}
We will solve (\ref{eq:master_K=J+2}) numerically and show the
superradiant instability.

\subsection{Stability analysis}\label{sec:KAK>0}
\subsubsection{A Method to Study the Stability }

We will find the instability by shooting method. 
Then, since the master equation~(\ref{eq:master_K=J+2}) is not self adjoint
form, we should put $\omega=\omega_R+i\omega_I$
($\omega_R,\omega_I\in\bm{R}$) and there are two shooting parameter, 
$\omega_R$ and $\omega_I$.
However, if the purpose is to find the onset of instability, 
the number of shooting parameter can be reduced to 
one~\cite{Press:1973,Kunduri:2006qa,Murata:2008yx}. 

We have separated the time dependence as 
$h_{\mu\nu}\propto e^{-i\omega t}$ in~(\ref{eq:h_K=J+2}). 
Therefore, unstable mode satisfies
$\text{Im}\,\omega>0$. Thus,   
the boundary condition for regularity at the horizon becomes
\begin{equation}
\Phi \rightarrow e^{- i\{\omega-2(J+2)\Omega_H\} r_\ast}
 \quad( r\rightarrow  r_+)\ .
\label{eq:MPbc1}
\end{equation}
Then, the general form of wave function at infinity becomes
\begin{equation}
 \Phi \rightarrow Z_1  r^{-5/2} + Z_2
  r^{3/2}
 \quad( r\rightarrow \infty)\ ,
\label{eq:MPbc2}
\end{equation}
where $Z_1,Z_2$ are constants. 
For regularity at infinity, the condition $Z_2=0$ must be satisfied. 
Therefore, the boundary conditions which unstable mode satisfies are
\begin{equation}
\Phi \rightarrow e^{- i\{\omega-2(J+2)\Omega_H\} r_\ast}
 \quad( r\rightarrow  r_+)\ ,\quad
\Phi \rightarrow 
Z_1 r^{-5/2}
 \quad( r\rightarrow \infty)\ .
\label{eq:MPbc}
\end{equation}

Now, we consider the marginally stable mode.
then, we can set $\text{Im}\,\omega=0$.
In this case of $\omega\in \bm{R}$, Wronskian of $\Phi_{K}$ is conserved, that is,
\begin{equation}
 \text{Im}\left[\Phi^\ast\frac{d}{d r_\ast}\Phi\right]^{ r= r_2}_{ r= r_1}
 = 0\ ,
\label{eq:Wronskian}
\end{equation}
for any $ r_1$ and $ r_2$. We take $ r_1= r_+$ and $ r_2=\infty$.  
Then, from Eq.~(\ref{eq:Wronskian}), we can get the relation,
\begin{equation}
 2(J+2)\Omega_H - \omega =
  -4L^{-2}\,\text{Im}(Z_1Z_2^\ast)\ .
\label{eq:in_out_rel}
\end{equation}
where we have used the asymptotic form of (\ref{eq:MPbc1}) and (\ref{eq:MPbc2}).
To avoid divergence at infinity, $Z_2 = 0$ must be satisfied. 
Then, we can get
\begin{equation}
 \omega = 2(J+2)\Omega_H\ .
\label{eq:AdS_omega_rel}
\end{equation}
Therefore, equation which we should solve is
\begin{equation}
-\frac{d^2 \Phi}{d r_\ast^2} + \hat{V}( r)\Phi = 0\ ,
\label{eq:master_AdS}
\end{equation}
where
\begin{equation}
 \hat{V}( r) \equiv V( r) - 4(J+2)^2(\Omega_H-\Omega(r))^2\ .
\label{eq:hatV}
\end{equation}
The boundary condition can be obtained by substituting
$\Phi=\Phi(r_+)+\Phi'(r_+)(r-r_+)$ into~(\ref{eq:hatV}) and it is given by
\begin{equation}
 \frac{\Phi'(r_+)}{\Phi(r_+)}=\frac{r_+^4+2\mu
  a^2}{r_+^4}\frac{V'(r_+)}{G'(r_+)^2}\ .
\label{eq:bc}
\end{equation}
There is only one shooting parameter, $\Omega_H$, in (\ref{eq:master_AdS}).


\subsubsection{Limit of Small Kerr-AdS Black
   Holes}\label{sebsec:SmallAdS}
Before the numerical calculation, it is important to solve master
equation~(\ref{eq:master_AdS}) analytically in some limit~\cite{Konoplya:2002zu}. It may be
useful to check the numerical calculation.  
We consider Kerr-AdS black holes in the limit of $r_+\rightarrow 0$. 
Then, master equation~(\ref{eq:master_K=J+2}) can
be solved exactly. The solution which approach to zero at infinity is
\begin{equation}
 \Phi = 
 \frac{ (r/L)^{7/2+2J}}{(1+r^2/L^2)^{J+3}}
 F\left((J+2)\Omega_HL+J+3,-(J+2)\Omega_HL+J+3;3;\frac{1}{1+r^2/L^2}\right)\
 ,
\label{eq:r_+=0}
\end{equation}
where $F(\alpha,\beta;\gamma;z)$ is Gauss hypergeometric function.
Then, the asymptotic form of $r\rightarrow 0$ becomes
\begin{multline}
 \Phi=
 \frac{2(2J+2)!}{\Gamma[(J+2)\Omega_HL+J+3]\Gamma[-(J+2)\Omega_HL+J+3]}
\left(\frac{r}{L}\right)^{-2J-5/2}\\
-\frac{4(-1)^{2J+3}}{(2J+3)!\Gamma[(J+2)\Omega_HL-J]\Gamma[-(J+2)\Omega_HL-J]}
\left(\frac{r}{L}\right)^{2J+7/2}\ln\left(\frac{r}{L}\right).
\label{eq:1}
\end{multline}
For the regularity at horizon, the first term of (\ref{eq:1}) must
vanish. 
Thus, we can get $\Omega_HL = (J+3+p)/(J+2)$ where $p=0,1,2,\cdots$. 
This calculation is to see the onset of the instability and the lowest value of
$\Omega_H$ is important. The lowest value of $\Omega_H$ is given by 
\begin{equation}
 \Omega_HL = \frac{J+3}{J+2}\ .
\label{eq:Omega_AdS}
\end{equation}
Numerical result must approach this value in the limit of
$r_+\rightarrow 0$. 

\subsubsection{Onset of superradiant instability}\label{sec:SRnumerical}
\begin{figure}
\begin{center}
\includegraphics[height=7cm,clip]{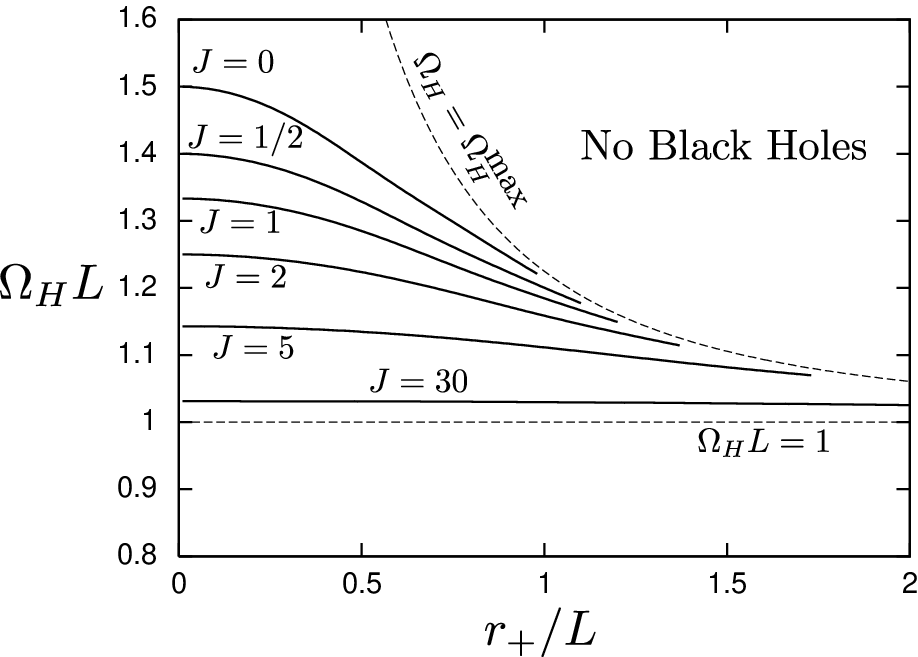}
\caption{\label{fig:OmegaH_r+}
The onset of superradiant instability is depicted in $\Omega_H$-$r_+$
 diagram. We plot onset lines for $J=0,1/2,1,2,5,30$ modes by solid
 lines. At the above dashed line, the Kerr-AdS$_5$ black holes become extreme
and, in the region of upper right, there is no black hole solutions.
The below dashed line is $\Omega_HL=1$.
We can see that onset line of higher
 $J$ modes appear at lower $\Omega_H$ and these lines approach
 $\Omega_HL=1$ for $J\rightarrow \infty$.
}
\end{center}
\end{figure}

Now, we shall solve~(\ref{eq:master_AdS}) numerically. 
Using the Lunge-Kutta algorithm, we integrate the
Eq.~(\ref{eq:master_AdS}) from the horizon to infinity with various $\Omega_H$. 
The boundary conditions at the horizon are given by~(\ref{eq:bc}). 
Then, the general form of the wave function at infinity is given by~(\ref{eq:MPbc2}).
We can see that, at some value of $\Omega_H$, the $Z_2$ flip the
sign. It means that $Z_2=0$ mode exists. We will search such 
$\Omega_H$ numerically and plot the result in $\Omega_H$-$r_+$
diagram. The result is in Figure.\ref{fig:OmegaH_r+}. 
The curves represent borderline of stability and
instability of each mode,
that is, each mode is stable below the curve, while they are unstable
above the curve. 
From this figure, we can read off that, in the limit of
$ r_+\rightarrow 0$, these curves for each mode approach 
$\Omega_H=(J+3)/(J+2)$. This result is consistent for analytical calculation
in section \ref{sebsec:SmallAdS}.
We can also see that, for higher $J$ mode, the instability occurs at
a lower angular velocity. 
These curves seem to approach $\Omega_HL=1$ for large $J$. 
These properties are the same for $D=7,9,11,\cdots$ cases~\cite{Kunduri:2006qa}.

It is surprising that these results have been already seen in dual
gauge theory~\cite{Murata:2008bg,Hawking:1999dp}. 
In~\cite{Murata:2008bg}, effective mass term for scalar fields of dual
 gauge theory have been obtained as
\begin{equation}
 m_\text{eff}^2=(2J+1)^2L^{-2}-4\Omega_H^2K^2 \ .
\end{equation}
Because of $|K|\leq J$, if $\Omega_HL<1$ satisfied, 
$m_\text{eff}^2$ is positive for any $J$ and $K$. 
However, if $\Omega_HL>1$, $m_\text{eff}^2$ can be negative for
large $J$ and $K$ modes. 
Thus, we see that, for $\Omega_HL>1$, dual gauge theory is unstable and 
higher $J$ mode becomes 
tachyonic first as $\Omega_H$ increases. 
These results are the same for superradiant instability of Kerr-AdS$_5$ black holes.

\section{Gregory-Laflamme instability of Kerr-AdS$_5\times S^5$
 spacetimes}
\label{sec:GL}

\subsection{Perturbation Equation}

In the previous section, we have seen the superradiant instability of
Kerr-AdS$_5\times S^5$ spacetime. 
The superradiant instability breaks the symmetry of Kerr-AdS$_5$. 
In this section, we will consider the Gregory-Laflamme instability of
Kerr-AdS$_5\times S^5$ spacetimes. This instability breaks the symmetry
of $S^5$. Thus, we must see the Kaluza-Klein modes of the perturbations which
have been neglected in the previous section. 

We will consider only the metric fluctuations on the AdS$_5$
part of the spacetime, that is,
\begin{equation}
\begin{split}
 &g'_{MN}dx^Mdx^N=g_{MN}dx^Mdx^N+h_{\mu\nu}(x^\mu)Y_\ell(\Omega_5)dx^\mu dx^\nu\ ,\\
 &\bm{F}'=2^{3/2}L^{-1} (\bm{\epsilon}'_{\text{AdS}_5}+\bm{\epsilon}_{S^5})
\end{split}
\label{eq:fluc}
\end{equation}
where $Y_\ell(\Omega_5)$ is spherical harmonics on $S^5$ which
satisfy
\begin{equation}
 \nabla_{S^5}^2Y_\ell=-\ell(\ell+4)Y_\ell\ ,
\label{eq:Y_ell_def}
\end{equation}
here $\nabla_{S^5}^2$ is the Laplacian of $S^5$ and $\ell=0,1,2,\cdots$.
The $\bm{\epsilon}'_{\text{AdS}_5}$ in (\ref{eq:fluc}) is the volume form of 
$g'_{\mu\nu}=g_{\mu\nu}+h_{\mu\nu}(x^\mu)Y_\ell(\Omega_5)$.
Since, in the case of Schwarzschild-AdS$_5\times S^5$,
the Gregory-Laflamme instability has been found in
these fluctuations~\cite{Hubeny:2002xn}, the instability of
Kerr-AdS$_5\times S^5$ must also appear in these fluctuations.
Here, we should notice that, in (\ref{eq:fluc}), $h_{\mu\nu}$ depends on
the coordinates on $S^5$. It is essential to see the Gregory-Laflamme
instability.
Then, from (\ref{eq:BGEOM1}), we can obtain the perturbation equations as
\begin{equation}
 \delta
  G_{\mu\nu}=\frac{6}{L^2}h_{\mu\nu}
-\frac{\varepsilon}{L^2}\left(h_{\mu\nu}-\frac{1}{2}h\right)\ ,
\label{eq:masssive}
\end{equation}
where $\varepsilon=\ell(\ell+4)/2$ and
$\delta G_{\mu\nu}$ is defined in~(\ref{eq:delG}).
From (\ref{eq:BGEOM2}), we can get 
\begin{equation}
 h(x^\mu) \partial_a Y_\ell(\Omega_5)=0\ ,
\label{eq:h=0}
\end{equation}
where $a$ is index on $S^5$.
In the case $\ell=0$, (\ref{eq:h=0}) is trivially satisfied and
(\ref{eq:masssive}) reduces to (\ref{eq:tensor_EOM}). 
For $\ell\geq 1$, (\ref{eq:h=0}) implies $h(x^\mu)=0$. 
Then, as a constraint equation of (\ref{eq:masssive}), we can get
transverse condition of $h_{\mu\nu}$ as shown in the appendix \ref{app:tt}. 
Thus, even for Kaluza-Klein modes, we can use transverse traceless
conditions, 
\begin{equation}
 \nabla^\nu h_{\mu\nu}=g^{\mu\nu}h_{\mu\nu}=0\ .
\label{eq:TT}
\end{equation}
To separate variables of equations~(\ref{eq:masssive}) and
(\ref{eq:TT}), we will use the formalism in the section \ref{sec:formalism}
again. 
We can expand $h_{\mu\nu}$ by the Wigner function $D^J_{KM}$ and
obtain ordinary differential equations labeled by $(J,K,M)$. 
We will not study the stability of all modes, but we will consider only 
$J=M=K=0$ mode. 
The Gregory-Laflamme instability of Schwarzschild-AdS$_5\times S^5$
spacetime was found in the 
s-wave of AdS$_5$~\cite{Hubeny:2002xn} and, 
thus, we can expect that, in the Kerr-AdS$_5\times S^5$ spacetime, 
the instability appears in most symmetric mode, $J=M=K=0$. 
The metric perturbation for this mode is given by
\begin{multline}
 h_{\mu\nu}(x^\mu)dx^\mu dx^\nu = 
h_{tt}(r) dt^2  + 2h_{t r}(r) dtd r + h_{ r r}(r) d r^2
+ 2h_{t3}(r) dt\sigma^3\\
 + 2h_{ r3}(r) d r\sigma^3 + 2h_{+-}(r) \sigma^+ \sigma^-
+ h_{33}(r) \sigma^3 \sigma^3\  ,
\label{eq:GLK0}
\end{multline}
where 
we assume that the metric perturbation $h_{\mu\nu}$ does not depend on $t$,
in order to see the onset of Gregory-Laflamme instability.
The stability of $\ell=0$ mode of this perturbation is shown in
Appendix~\ref{app:K=0} and, thus, we will consider $\ell=1,2,3,\cdots$.
We substitute (\ref{eq:GLK0}) into (\ref{eq:masssive}).
Then, from $tr$ and $r3$ components of the perturbation
equation, we can obtain
\begin{equation}
h_{tr}=h_{r3}=0\ .
\end{equation}
Now, we introduce dimensionless variables, $\alpha,\beta,\delta,\eta$ and
$\zeta$, as
\begin{equation}
\begin{split}
&h_{tt}=-\left(1+\frac{r^2}{L^2}-\frac{2 \mu}{r^2}\right) \alpha\ ,\quad
h_{rr}=\frac{\beta}{G(r)}\ ,\quad
h_{+-}=\frac{r^2}{4}\delta\\
&h_{33}=\frac{r^2}{4}\left(1+\frac{2\mu a^2}{r^4}\right) \eta\ ,\quad
h_{t3}=\frac{2 \mu a}{r^2} \zeta\ .
\end{split}
\end{equation}
Hereafter, we put $L=1$ to simplify expressions.
Then, the traceless condition $h=0$ can be written as
\begin{multline}
(r^4+r^2-2\mu)(r^4+2\mu a^2)\alpha
+r^6 G(r)\beta
+r^6 G(r)\delta\\
+(r^4+r^2-2\mu)(r^4+2\mu a^2)\eta
+16\mu^2 a^2\zeta=0\ ,
\label{eq;traceless}
\end{multline}
and the $r$-component of transverse condition $\nabla^\nu h_{\mu\nu}=0$ is
\begin{equation}
\begin{split}
&-(r^4+r^2-2\mu)\{r^8+2\mu(1+2a^2)r^4-4\mu^2a^2(1+a^2)\}\alpha\\
&+r^9G(r)^2\beta'
+r^4G(r)\{4r^6+3r^4-4\mu(1-a^2)r^2+2\mu a^2\}\beta
+r^8 G(r)^2\delta\\
&-(r^4+2\mu a^2)
\{r^8+2r^6-(2\mu a^2+4\mu-1)r^4-4\mu(a^2+1)r^2+2\mu(2\mu+2\mu a^2+a^2)\} \eta\\
&-16\mu^2a^2\{3r^4+2r^2-2\mu(1-a^2)\}\zeta
=0\ ,
\label{eq:trans}
\end{split}
\end{equation}
where $'\equiv d/dr$.
We can see that, using (\ref{eq;traceless}) and (\ref{eq:trans}),
$\alpha$ and $\zeta$ can be eliminated.\footnote{
We can eliminate other variables, such as ($\alpha,\delta$), ($\delta,\eta$).
However, if we eliminate these variables, the singular point will appear
at $r_+<r<\infty$ in the resultant equations~\cite{Hubeny:2002xn}. 
The variables ($\alpha,\zeta$) are the best variables for elimination as
far as we can see.
}
Then, $rr$,$+-$ and $33$ component of (\ref{eq:masssive}) are
given by
\begin{align}
&-r^6 G(r) \beta''
-r\{13r^6+9r^4-10\mu(1-a^2)r^2+2\mu a^2\} \beta' \nonumber\\
&-2\{(16-\varepsilon)r^6+6r^4-4\mu a^2\}\beta
+4(r^4-2\mu a^2)\delta
+4(r^4+2\mu a^2)\eta
=0\ , \label{eq:rr}\\
\nonumber\\
&4r^5G(r) \beta'
+8(2r^6+r^4-2\mu a^2)\beta
-r^6G(r)\delta'' \nonumber\\\
&-r\{5r^6+3r^4-2\mu(1-a^2)r^2-2\mu a^2\} \delta'
+2(\varepsilon r^6+8\mu a^2) \delta
-8(r^4+2\mu a^2)\eta
=0\ , \label{eq:+-}\\
\nonumber\\\
&2r^5G(r)(r^4+2\mu a^2) \beta''
+2(r^4+2\mu a^2)\{11r^6+7r^4-6\mu(1-a^2)r^2-2\mu a^2\}\beta' \nonumber\\\
&+16r^3(3r^2+1)(r^4+2\mu a^2) \beta
-2(r^8-4\mu r^6+4\mu a^2r^4+4\mu^2a^4)\delta'
-8r^3(r^4+2\mu a^2)\delta \nonumber\\\
&-r^5G(r)(r^4+2\mu a^2)\eta''
-r^4G(r)(5r^4-6\mu a^2) \eta'
+2\varepsilon r^5(r^4+2\mu a^2) \eta
=0\ . \label{eq:33}
\end{align}
We can check that 
the other components of (\ref{eq:masssive}) are derived from 
(\ref{eq:rr}), (\ref{eq:+-}) and (\ref{eq:33}). 
There are three degree of freedom in $J=M=K=0$ 
mode.\footnote{
In the case $\ell=0$, the gauge freedom are restored 
and degree of freedom becomes one 
as explained in appendix \ref{app:K=0}.}

\subsection{Onset of the Gregory-Laflamme instability}
We will solve equations (\ref{eq:rr}), (\ref{eq:+-}) and (\ref{eq:33})
numerically and see the onset of the Gregory-Laflamme instability.
First, we derive the boundary conditions at the horizon.
Substituting 
\begin{equation}
 \beta=b_0+b_1(r-r_+)\ ,\quad
 \delta=d_0+d_1(r-r_+)\ ,\quad
 \eta=e_0+e_1(r-r_+)\ ,\quad
\end{equation}
into (\ref{eq:rr}), (\ref{eq:+-}) and (\ref{eq:33}),
we obtain
\begin{equation}
\begin{split}
&b_1=\frac{(\varepsilon r_+^6 -16r_+^6+r_+^4\varepsilon -24r_+^4-8r_+^2-4\mu)b_0
+4(r_+^4+r_+^2-2\mu)d_0}
{4r_+(r_+^6+2r_+^4+r_+^2-\mu)}\ ,\\
&d_1 =\frac{8(r_+^6+2r_+^4+r_+^2+\mu)b_0
+(\varepsilon r_+^6+\varepsilon r_+^4-4r_+^4+16\mu-4r_+^2)d_0}
{2r_+(r_+^6+2r_+^4+r_+^2-\mu)}\ ,\\
&e_0 = -2b_0-d_0\ .
\end{split}
\end{equation}
Free parameters $b_0,d_0,e_1$ remain. 
However, we can set $e_1=1$ by the rescale of $\beta,\delta,\eta$. 
Hence, parameters which we should set at horizon are $b_0,d_0$.
On the other hand, at $r\rightarrow \infty$, the
growing mode of $\beta,\delta,\eta$ becomes
\begin{equation}
 \delta\simeq C_1\,r^\ell\ ,\quad
\eta\simeq C_2\,r^\ell\ ,\quad
\beta\simeq \frac{C_1+C_2}{\ell+3}\,r^{\ell-2}+C_3r^{\ell-4}\ .
\end{equation}
Thus, for large $r$, we can get the coefficients of the growing modes approximately as
\begin{equation}
 C_1=\delta/ r^\ell\ ,\quad
 C_2=\eta/ r^\ell\ ,\quad
 C_3=\left(\beta-\frac{C_1+C_2}{\ell+3}\,r^{\ell-2}\right)/ r^{\ell-4}\ .
\end{equation}
These $C_1,C_2,C_3$ must be zero at infinity\footnote{
For $\ell=1,2,3,4$, $\beta$ is not singular even if $C_3\neq 0$, but for
the regularity of $\zeta$, we need $C_3=0$.
}. 

\begin{figure}
\begin{center}
\includegraphics[height=7cm,clip]{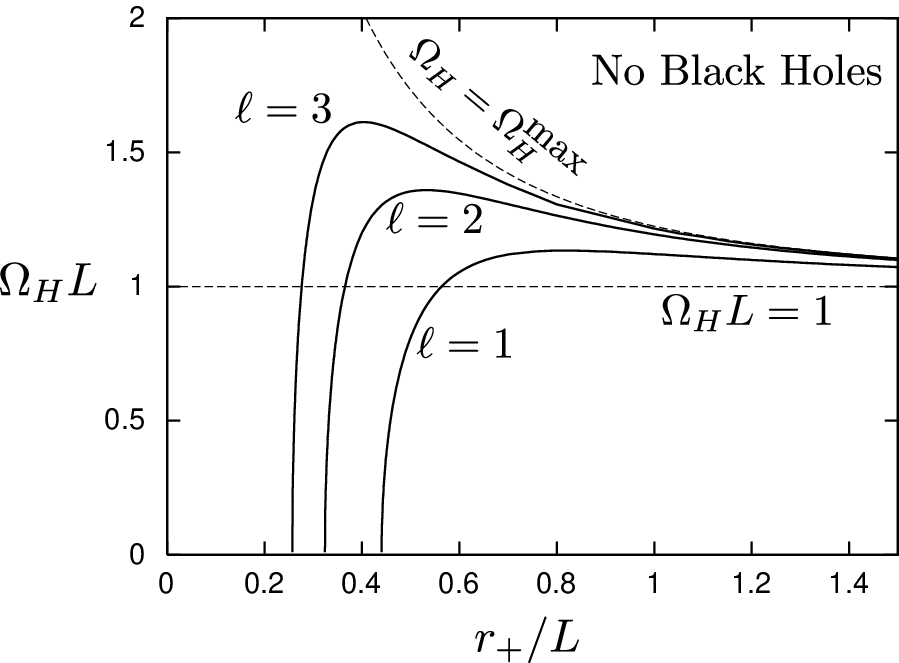}
\caption{\label{fig:GL}The onset lines of
Gregory-Laflamme instability are depicted.
We depicted them for $\ell=1,2,3$ by solid lines. 
Above regions of each line, Kerr-AdS$_5$ is unstable against Gregory-Laflamme
instabilities of each mode.
These lines have maximum
value and approach $\Omega_HL=1$ for $r_+\rightarrow\infty$.
At the above dashed line, the Kerr-AdS$_5$ black holes become extreme
and, in the region of upper right, there are no black hole solutions.
The below dashed line is $\Omega_HL=1$, which is the onset of the superradiant
instability. 
}
\end{center}
\end{figure}

Now, we can start the numerical integration. 
We solve (\ref{eq:rr}), (\ref{eq:+-}), (\ref{eq:33})
from $r_1=r_++1.0\times 10^{-5}L$ to $r_2=1.0\times 10^4L$ by the Runge-Kutta
algorithm. Input parameters are $r_+,\Omega_H,b_0,d_0$ and 
we can get 
$C_i=C_i(r_+,\Omega_H,b_0,d_0)$ $(i=1,2,3)$ at $r=r_2$. 
For fixed $r_+$, we look for $b_0,d_0,\Omega_H$ which satisfy $C_i=0$
by the Newton-Raphson method. 
We repeat this procedure with various $r_+$.
The result is given in Figure.~\ref{fig:GL}. 
These lines have maximum
value and approach $\Omega_HL=1$ for $r_+\rightarrow\infty$.
We can see that $\ell=1$ mode is
relevant for the onset of the Gregory-Laflamme instability. 
It is remarkable that onset line of Gregory-Laflamme and superradiant
instabilities intersect each other. Thus, both of the instabilities can
appear in Kerr-AdS$_5\times S^5$ spacetimes.
In the limit of $\Omega_H\rightarrow 0$, 
we can read off the onset of the instability as 
$r_+/L=0.4402(\ell=1),0.3238(\ell=2),0.2570(\ell=3)$.
It is consistent with the instability of 
Schwarzschild-AdS$_5\times S^5$ spacetimes~\cite{Hubeny:2002xn}.

\section{Phase structure}\label{sec:Phase}
In this section, taking into account superradiant and
Gregory-Laflamme instabilities, the phase
structure of the Kerr-AdS$_5\times S^5$ spacetime is revealed.
Here, we need some comments on the stability analyses performed in
section \ref{sec:SR} and \ref{sec:GL}. 
We studied some specific modes and found instabilities.
However, we did not study all modes of perturbations and, 
strictly speaking, onsets of superradiant and Gregory-Laflamme
instabilities can be changed by all modes analysis.
To make sure that the  onset of instabilities that we derived gives a true onset
of instability, we studied the stability of $(J,M,K=J+2)$, $\ell\neq 0$
modes in Appendix~\ref{app:K=J+2_massive}. 
As the result, we found the instability whose onset is given by
$\Omega_HL\simeq (J+3+\ell/2)/(J+2)$. This result suggest that 
mass term of graviton lifts up the onset of instability
and these are not relevant to see the onset of superradiant instability.
The result of Appendix~\ref{app:K=J+2_massive} also suggests that
the Gregory-Laflamme type instability is not found in 
higher modes in AdS$_5$ part of spacetime.
Because of above reason, we regard the result derived in section \ref{sec:SR}
and \ref{sec:GL} as a true onset of instabilities and
reveal the phase structure
of Kerr-AdS$_5\times S^5$ spacetimes. 

Until previous section, 
we have been using parameters $(r_+,\Omega_H)$. However, 
for comparison with the gauge theory, the horizon radius $r_+$ is not a good parameter
because $r_+$ is not defined in the gauge theory.
Therefore, we see the phase diagram by
thermodynamical parameters, the temperature $T$ and angular velocity
$\Omega_H$.
The temperature is defined by
\begin{equation}
 T=
\frac{2(1-\Omega_H^2L^2)r_+^2/L^2+1}{2\pi r_+}
\sqrt{\frac{1+r_+^2/L^2}{(1-\Omega_H^2L^2)r_+^2/L^2+1}}\ .
\label{eq:T}
\end{equation}
Using this equation, we can map Figure.\ref{fig:GL} onto $T$-$\Omega_H$
diagram. The result is given in Figure.\ref{fig:phase}.
The solid and dashed lines are the onset of Gregory-Laflamme and
superradiant instabilities, respectively. These lines cross each other
and we can see five phases in this diagram. 
In the ``Stable'' region, Kerr-AdS black holes are stable. 
In the ``SR'' and ``GL'' region, black holes are
unstable against superradiant and Gregory-Laflamme instabilities,
respectively. In ``SR{\&}GL'' region, black holes are unstable
against both of them. In ``No Black Holes'' region, there is no black
hole solution. 

\begin{figure}
  \begin{center}
      \includegraphics[height=6cm,clip]{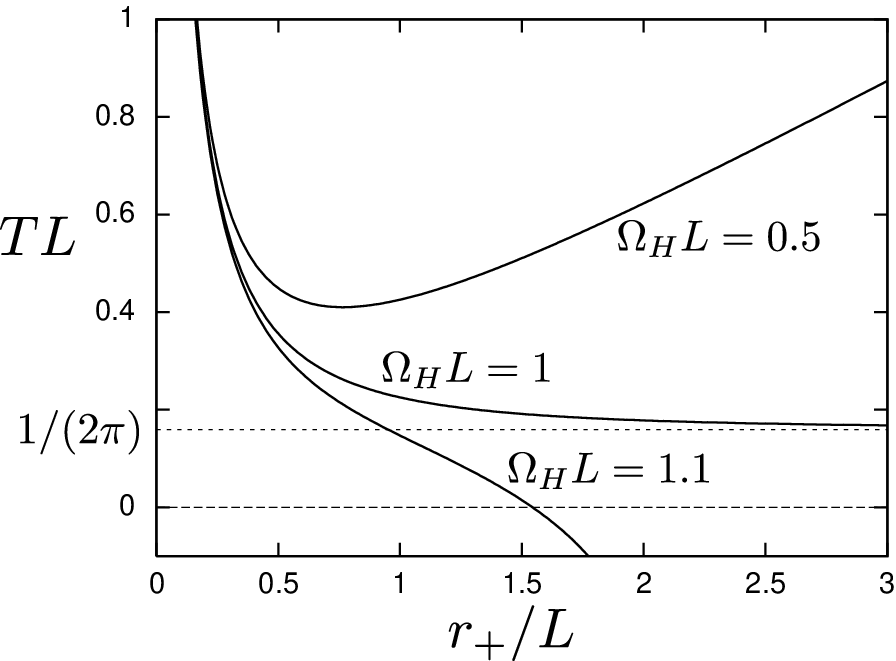}
    \caption{\label{fig:T_rp}We plot the temperature of Kerr-AdS$_5$
   black holes as a function of $r_+$ for $\Omega_HL=0.5,1,1.1$.
For $\Omega_HL<1$, there are two $r_+$ giving the same temperature and
   the temperature $T$ has a minimal value. 
For $\Omega_HL=1$, the temperature monotonically decreases and approaches
   $TL=1/(2\pi)$. 
For $\Omega_H>1$, the temperature becomes zero at some value of $r_+$.
}
  \end{center}
\end{figure}

To get this phase diagram, we need some attentions. 
In Figure.\ref{fig:T_rp}, we plot the temperature as a function of
$r_+$. In the case of $\Omega_HL<1$, there are two $r_+$ giving the same
temperature. We will call these phases as small and large black hole
phases. There is no one-to-one correspondence for $r_+$ and
$T$. Since the Gregory-Laflamme instability appears in the small black hole
phase, we chose the small black hole phase to depict the phase diagram.
We can also see that the temperature has minimal value 
$T_\text{min}(\Omega_H)>0$. The ``No Black Holes'' phase in
Figure.\ref{fig:phase} comes from this bound.
In the case of $\Omega_HL=1$, the temperature
$T$ has no minimal value, but it is bounded by $T>1/(2\pi)$. 
For $\Omega_HL>1$, $T$ becomes zero at some value of $r_+$. Thus, ``No Black
Holes'' phase is vanishing for $\Omega_HL>1$ in Figure.\ref{fig:phase}. 
The $\Omega_H=\Omega_H^\text{max}$ line in Figure.\ref{fig:GL}, 
has been mapped onto a ray of $T=0$ and $\Omega_H>1$.

\section{Conclusions and Discussions}\label{sec:conc}

We have studied gravitational perturbations of Kerr-AdS$_5\times S^5$ spacetimes
with equal angular momenta. 
First, we studied the stability of Kerr-AdS$_5$ neglecting Kaluza-Klein
modes and found the superradiant instability. 
We could see that the
onset is given by $\Omega_HL=1$.
We also studied the stability including Kaluza-Klein modes of $S^5$ 
and found the Gregory-Laflamme instability. 
From these results, we made out the phase diagram of 
Kerr-AdS$_5\times S^5$ spacetime in Figure.\ref{fig:phase} and found
five phases in this diagram. 


It is surprising that superradiant and Gregory-Laflamme
instabilities can be understood by the dual gauge theory.
In the dual gauge theory, the angular
velocity of the horizon $\Omega_H$ is regarded as a chemical 
potential~\cite{Hawking:1998kw,Hawking:1999dp,Murata:2008bg}. 
It has been shown that, for $\Omega_HL>1$, the gauge theory is unstable and 
higher $J$ mode becomes tachyonic first as $\Omega_H$
increases~\cite{Murata:2008bg}. 
In gravity theory, we found same property, that is, 
we can see that, from Figure.\ref{fig:OmegaH_r+}, higher $J$ mode becomes 
unstable first as $\Omega_H$ increases.
This remarkable coincidence of gravity and gauge
theories gives a strong evidence for the AdS/CFT correspondence. 

There are also several works on the Gregory-Laflamme instability from
the gauge
theory point of view~\cite{Aharony:2004ig,Harmark:2004ws,Hanada:2007wn,Hollowood:2006xb}. 
Especially, in~\cite{Hollowood:2006xb}, 
$\mathcal{N} =4$ SYM on $S^3$, which is dual to
the Schwarzschild-AdS$_5\times S^5$ spacetime, is studied with weak 't~Hooft coupling. 
In their work, for high temperature, they found a new saddle point in which $SO(6)$ R-symmetry
is spontaneously broken and $SO(5)$ symmetry is remained. 
The $SO(6)$ R-symmetry in the gauge theory corresponds to the symmetry of
internal space $S^5$ in the dual gravity theory and, thus, this appearance
of the new saddle point was identified to the Gregory-Laflamme
instability. 
For the Kerr-AdS$_5\times S^5$ spacetime, 
our result of Figure.\ref{fig:phase} gives a
prediction for the phase structure of the gauge theory.
However, unfortunately, there is no work for the Gregory-Laflamme
instability in the view of gauge theory for the rotating black hole.
It is interesting to extend the work of~\cite{Hollowood:2006xb} to
Kerr-AdS$_5\times S^5$ spacetime and compare with our result of
Figure.\ref{fig:phase}.

In the dual gauge theory, we can still introduce R-symmetry chemical
potentials. 
This theory corresponds to the R-charged black hole solution obtained
in~\cite{Cvetic:1999ne}.
For highly R-charged black holes, thermodynamical instability was
found~\cite{Cvetic:1999ne,Yamada:2006rx,Murata:2008bg}
and this instability can be understood by the dual gauge
theory~\cite{Yamada:2006rx,Murata:2008bg}.
However, the dynamical instability is not found through gravitational
perturbations. 
In the gauge theory, this instability is described by the appearance of
a tachyonic mode of scalar fields and, thus,  
this instability breaks the R-symmetry, $SO(6)$. 
On the other hand, in gravity theory,
$SO(6)$ symmetry comes from internal space $S^5$ and, 
therefore, we can expect that
this instability appears in Kaluza-Klein modes of $S^5$. 
The Kaluza-Klein modes can be regarded as a charged
field in the effective theory in AdS$_5$. In the system of a charged black hole and
charged field, superradiance occurs and superradiant instability is
caused in the AdS spacetime.
It is challenging to study the stability of R-charged black hole
taking into account the Kaluza-Klein modes of $S^5$.

In this paper, because of the practical reason, 
we could not study stability of Kerr-AdS$_5$ spacetimes
with independent angular momenta.
However, for Kerr-AdS$_5$ with one rotation, we may be able to find
a new kind of instability. 
In the case of asymptotically flat spacetimes, it was suggest that there is 
a phase transition between five-dimensional Kerr black hole and black
ring solutions~\cite{Emparan:2001wn}. 
For asymptotically AdS spacetimes, perturbative solution of black ring
has been found~\cite{Caldarelli:2008pz} and there may be a
transition between Kerr-AdS$_5$ black hole and the AdS black ring.


\vspace{1cm}
\centerline{\bf Acknowledgments}
We are grateful to J. Soda and R. A. Konoplya for
careful reading of this paper.
We would also like to thank Veronika E. Hubeny for telling us
numerical method to find the Gregory-Laflamme instability of
Schwarzschild-AdS$_5\times S^5$ spacetime.
The work is supported by the JSPS Grant-in-Aid for Scientific Research
No. 19 $\cdot$ 3715.

\appendix
\section{Kerr-AdS black hole with independent angular momenta}
\label{app:KerrAdS_indep}

The Kerr-AdS$_5$ spacetime with independent angular momenta is given 
by~\cite{Hawking:1998kw,Gibbons:2004js,Gibbons:2004uw,Gibbons:2004ai}
\begin{equation}
\begin{split}
ds^2 &= - \frac{\Delta}{\rho^2} \left(dt + \frac{a_1
      \sin^2\theta_1}{\Xi_1}d\phi_1 +
  \frac{a_2 \cos^2\theta_1}{\Xi_2} d\phi_2\right)^2 +
  \frac{\Delta_{\theta}\sin^2\theta_1}{\rho^2} \left(a_1 dt +
  \frac{(\bar{r}^2+a_1^2)}{\Xi_1} d\phi_1\right)^2 \\
  & + \frac{\Delta_{\theta_1}\cos^2\theta_1}{\rho^2} \left(a_2 dt +
  \frac{(\bar{r}^2+a_2^2)}{\Xi_2} d\phi_2\right)^2 + \frac{\rho^2}{\Delta} d\bar{r}^2 +
  \frac{\rho^2}{\Delta_{\theta_1}} d\theta_1^2 
\\
  & + \frac{(1+\bar{r}^2/L^2)}{\bar{r}^2 \rho^2}
  \left ( a_1a_2 dt + \frac{a_2 (\bar{r}^2+a_1^2) \sin^2\theta_1}{\Xi_1}d\phi_1
    + \frac{a_1 (\bar{r}^2 + a_2^2) \cos^2 \theta_1}{\Xi_2} d\phi_2
 \right)^2\ ,
\end{split}
\end{equation}
where
\begin{equation}
\begin{split}
  \Delta &= \frac{1}{\bar{r}^2} (\bar{r}^2 + a_1^2) (\bar{r}^2 + a_2^2) (1 + \bar{r}^2/L^2) - 2m\ ,\\
  \Delta_{\theta} &=  1 - a_1^2L^{-2} \cos^2\theta_1 - a_2^2 L^{-2}
    \sin^2\theta_1\ , \\
  \rho^2 &=  \bar{r}^2 + a_1^2 \cos^2\theta_1 + a_2^2 \sin^2\theta_1\ ,\\
  \Xi_i &= 1-a_i^2/L^2\ (i=1,2)\ ,
\end{split}
\end{equation}
This spacetime has symmetry $R_t\times U(1)^2$ generated by
$\partial_t$, $\partial_{\phi_1}$ and $\partial_{\phi_2}$.
Now, we consider Kerr-AdS spacetime with equal angular momenta,
$a_1=a_2\equiv a$. We introduce
new coordinates and parameter,
\begin{equation}
\begin{split}
&\theta=2\theta_1\ ,\quad
\phi=\phi_2-\phi_1\ ,\quad\psi=\phi_1+\phi_2-2at\ ,\\
&r^2=(\bar{r}^2+a^2)/(1-a^2)\ ,\quad
\mu=m/(1-a^2)^3\ .
\end{split}
\end{equation}
As a result, we can get the AdS part of~(\ref{eq:KerrAdS}). 

\section{Stability analysis for other modes}\label{app:other}
In section \ref{sec:SR} and \ref{sec:GL},
we studied specific modes and found superradiant and Gregory-Laflamme
instabilities. However, Gregory-Laflamme
instabilities may be changed by all modes analysis.
In this appendix, we will study the stability of other modes
which can be reduced to a single master equation.
These are $(J=0,M=0,K=0,1)$ with $\ell=0$ modes and
$(J,M,K=J+2)$ with any $\ell$ modes.
As the result of stability analysis, 
we can see some evidence that the onset of instabilities
derived in sections \ref{sec:SR} and \ref{sec:GL} gives a true onset of
instability.

\subsection{$(J=0,M=0,K=0)$ with $\ell=0$ mode} 
\label{app:K=0}
In perturbation equation~(\ref{eq:tensor_EOM}),
$(J=0,M=0,K=0,1)$ and $(J,M,K=J+2)$ modes can be reduce to a single master
equations. The stability of $(J,M,K=J+2)$ modes are studied in
section \ref{sec:SR}. Here, we shall consider $(J=0,M=0,K=0)$ mode.\footnote{
This mode have been already studied using other formalism~\cite{Bizon:2007zf}.}

As we have seen in section~\ref{sec:formalism}, there exist
$h_{tt},h_{tr},h_{rr},h_{t3},h_{r3},h_{+-},h_{33}$ fields in this mode. 
We set $h_{\mu\nu}$ as
\begin{multline}
 h_{\mu\nu}dx^\mu dx^\nu = e^{-i\omega t}\big[h_{tt}(r) dt^2  + 2h_{t r}(r) dtd r + h_{ r r}(r) d r^2
+ 2h_{t3}(r) dt\sigma^3\\
 + 2h_{ r3}(r) d r\sigma^3 + 2h_{+-}(r) \sigma^+ \sigma^-
+ h_{33}(r) \sigma^3 \sigma^3\big]\ .
\label{eq:h_K=0}
\end{multline}
With the gauge parameters 
\begin{equation}
\begin{split}
 \xi_A(x^\mu) =  \xi_A(r) e^{-i\omega t}\ ,\quad
 \xi_i(x^\mu) =  \xi_3(r) e^{-i\omega t}\sigma^3_i \ ,
\end{split} 
\end{equation} 
the gauge transformations $\delta h_{\mu\nu}= \nabla_\mu \xi_\nu + \nabla_\nu \xi_\mu$
for these components are given by
\begin{equation}
\begin{split}
%
&\delta h_{tt}=
-2i\omega \xi_t
-\frac{4\mu G(r)}{r^3} \xi_r\ , \quad
\delta h_{tr}=
\xi_t'
-\frac{4\mu}{r^3G(r)}\xi_t
-i\omega \xi_r
+\frac{8\mu}{r^5G(r)} \xi_3\ , \\
&\delta h_{t3}=
-\frac{2G(r)\mu a}{r^3} \xi_r
-i\omega \xi_3\ , \quad
\delta h_{rr}=
2 \xi_r'
+\frac{4\mu(r^2-2a^2)}{r^5G(r)} \xi_r\ , \\
&\delta h_{r3}=
-\frac{4\mu a}{r^3G(r)} \xi_t
+ \xi_3'
-\frac{2(r^4-2\mu r^2-2\mu a^2)}{r^5G(r)} \xi_3'\ , \\
&\delta h_{+-}=
rG(r) \xi_r \ ,\quad
\delta h_{33}=
\frac{G(r)(r^4-2\mu a^2)}{2r^3} \xi_r\ .
\end{split}
\label{eq:gauge0}
\end{equation}
Our gauge choices are
\begin{equation}
h_{tt}=h_{t3}=h_{33}=0\ .
\label{eq:K=0_gauge_choice}
\end{equation}
One can check that these are complete gauge fixing
from~(\ref{eq:gauge0}). 
After the gauge fixing, four fields $h_{t r},h_{ r r},h_{ r3},h_{+-}$
remain. However, all of them do not have degree of freedom. 
Substituting Eq.~(\ref{eq:h_K=0}) and Eq.~(\ref{eq:K=0_gauge_choice})
into Eq.~(\ref{eq:tensor_EOM}), we can get three constraint
equations and one degree of freedom remains. 
Therefore, we can get a single master equation. 
The equation can be written in the
Schr\"{o}dinger form as\footnote{
The detail calculations are very similar to~\cite{Murata:2008yx} and 
we have omitted the most of them.
}
\begin{equation}
 -\frac{d^2\Phi_0}{d r_\ast^2} + V_0( r)\Phi_0 = \omega^2 \Phi_0
\label{eq:K=0_master}
\end{equation}
where
\begin{equation}
 \Phi_0\equiv 
\frac {( r^4-2\mu a^2)({ r}^{4}+2\mu a^2)^{1/4}}
{ r^{3/2}(3 r^4+2\mu a^2)}h_{+-}\ ,
\end{equation}
and the tortoise coordinate $r_\ast$ is defined in~(\ref{eq:kame}).
The potential $V_0(r)$ is determined by
\begin{equation}
\begin{split}
V_0( r) =&\,
 \frac{G(r)}{4(3 r^4+2\mu a^2)^2( r^4+2\mu a^2)^3 r^2}\\
\times& \big[
135 r^{22}/L^2
+315 r^{20}
+18\mu(9+43a^2/L^2) r^{18}
+2430\mu a^2  r^{16}\\
&+8\mu^2 a^2 (174+55a^2/L^2) r^{14}
+5400\mu^2 a^4  r^{12}
+16 \mu^3 a^4(363-193a^2/L^2) r^{10}\\
&+2608\mu^3 a^6  r^8
+80\mu^4 a^6 (76-49a^2/L^2) r^6
-2064\mu^4 a^8  r^4\\
&+32\mu^5 a^8 (1-a^2/L^2) r^2
-160\mu^5 a^{10}
\big]\ .
\end{split}
\label{eq:K=0pot}
\end{equation}

\begin{figure}[h]
  \leavevmode
  \begin{center}
      \includegraphics[height=5cm,clip]{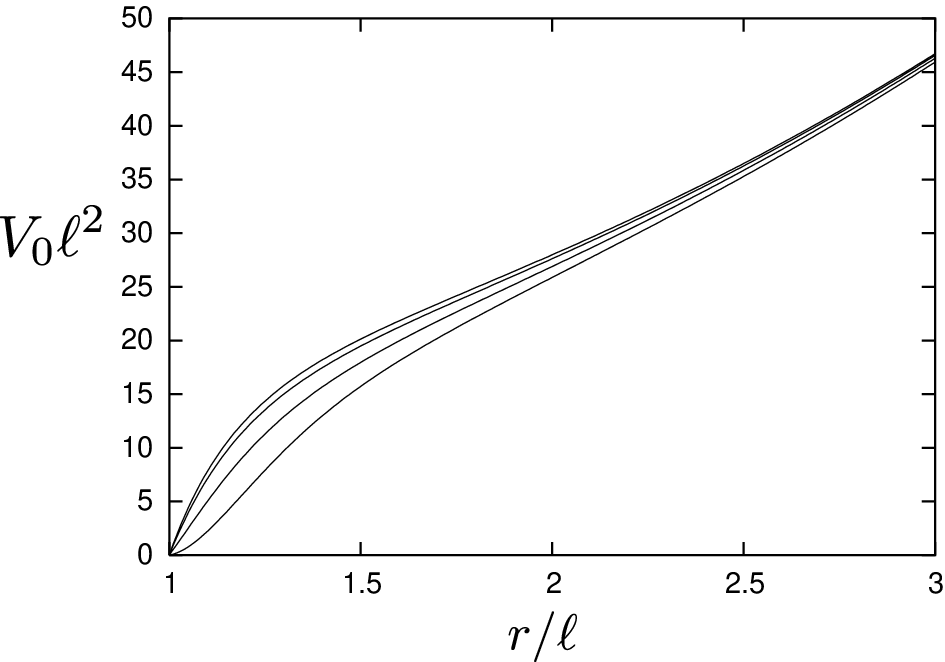}
    \caption{\label{fig:potential_V0_AdS}Typical profiles for the potential
 $V_0$ are depicted. 
   We put $r_+=1.0L$. From top to
 bottom, each curve represents the potential for 
 $\Omega_H/\Omega_H^\text{max}=0.1,0.7,0.9,0.99$. We see the positivity
 of these potentials.
}
  \end{center}
\end{figure}

We consider the stability of this $J=M=K=0$ mode. 
In this mode, the master
equation~(\ref{eq:K=0_master}) is in the Schr\"{o}dinger form. Therefore,
positivity of $V_0$ means stability of this mode.
The typical profile of $V_0$ is shown in
Figure.~\ref{fig:potential_V0_AdS}. We can see the positivity of this
potential from this figure. In fact, the positivity can be checked from
the expression~(\ref{eq:K=0pot}). From Eq.~(\ref{eq:Omegamax}) and
(\ref{eq:mu_a_to_r+_Omega}), we obtain
\begin{equation}
 a^2\leq
  \frac{r_+^4}{(1+r_+^2/L^2)^2}\left(\frac{1}{2r_+^2}+\frac{1}{L^2}\right)\ .
\end{equation}
The right hand side is an increasing function of $r_+$ which approaches to
$L^2$ in the limit of $r_+\rightarrow \infty$. Thus, we can get
the inequality $a^2\leq L^2$. Therefore, negative terms in the big
brackets of Eq.~(\ref{eq:K=0pot}) are $r^4$ and $r^0$ terms. 
To see positivity of $V_0(r)$, we focus on $r^6$, $r^4$ and $r^0$ terms
in the big bracket of Eq.~(\ref{eq:K=0pot}).
After dividing them by $16 \mu^4 a^6$, these terms become
\begin{equation}
 f(r)=5 (76-49a^2/L^2) r^6
-129 a^2  r^4
-10\mu a^4\ .
\label{eq:g}
\end{equation}
If $f(r)$ is positive, $V_0(r)$ is also positive. 
Now, we substitute Eq.~(\ref{eq:mu_a_to_r+_Omega}) into
Eq.~(\ref{eq:g}). 
Because of $\Omega\leq\Omega_H^\text{max}$ and $r\geq r_+$, 
we can put $\Omega_H=s^2/(1+s^2)\Omega_H^\text{max}$ for $s\geq 0$ and
$r^2=x^2+r_+^2$. Then, we can obtain
\begin{equation}
\begin{split}
f(r)=&\frac{L^6}{2(1+s^2)^2\alpha^2}\big[
\{760\alpha^2+1520\alpha^2s^2
+(760+1275\beta+270\beta^2)s^4\}x^6\\
&+\beta\{2280\alpha^2+4560\alpha^2s^2+3(717+1189\beta+270\beta^2)s^4\}x^4\\
&+\beta^2\{2280\alpha^2+4560\alpha^2s^2+3(674+1103\beta+270\beta^2)s^4\}x^2\\
&+\beta^3\gamma
\{1520\alpha^3+6080\alpha^3s^2+2\alpha(4051+7477\beta+3310\beta^2)s^4\\
&+4\alpha(1011+1397\beta+270\beta^2)s^6
+(626+997\beta+250\beta^2)s^8
\}
\big]\ ,
\end{split}
\end{equation}
where $\alpha=1+r_+^2/L^2$, $\beta=r_+^2/L^2$ and 
$\gamma=1/(2\alpha+4s^2\alpha+s^4)$.  We can see $f(r)\geq0$ explicitly. 
It means the stability of $J=M=K=0$ mode. 

\subsection{$(J=0,M=0,K=1)$ with $\ell=0$ mode}\label{app:K=1}
In $(J=0,M=0,K=1)$ mode, there are
$h_{t+},h_{r+},h_{+3}$ fields. 
We set $h_{\mu\nu}$ as
\begin{equation}
 h_{\mu\nu}dx^\mu dx^\nu = e^{-i\omega t}\big[2h_{t+}(r) dt\,\sigma^+ 
+ 2h_{ r+}(r) d r\,\sigma^+ 
+ 2 h_{+3}(r) \sigma^+ \sigma^3\big]\ ,
\label{eq:h_K=1}
\end{equation}
With the gauge parameter
$\xi_i(x^\mu)=e^{-i\omega t}\xi_+(r)\sigma^+_i$, 
the gauge transformations for these components are given by
\begin{equation}
\delta h_{t+} =
 -i\omega \xi_+ + \frac{4i\mu a}{ r^4}\xi_+\ ,\quad
\delta h_{ r+} =
\xi_+'-\frac{2}{ r}\xi_+\ ,\quad
\delta h_{+3} =
\frac{2i\mu a^2}{ r^4}\xi_+\ .
\end{equation}
Our gauge choice is 
\begin{equation}
 h_{+3}=0\ .
\label{eq:K=1_gauge}
\end{equation}
This condition fixes the gauge completely. 
After the gauge fixing, 
two fields $h_{t+}$ and $h_ {r+}$ are remained. However, because
one constraint exists in this mode, the physical degree of freedom
becomes one. Therefore, we can get one master equation. 
Substituting Eq.~(\ref{eq:h_K=1}) and Eq.~(\ref{eq:K=1_gauge}) into
Eq.~(\ref{eq:tensor_EOM}),
we can get the master equation for this mode,
\begin{equation}
 -\frac{d^2 \Phi_1}{d r_\ast^2} + V_1( r)\Phi_1 =
  [\omega-2\Omega_1( r)]^2 \Phi_1\ ,
\label{eq:master_K=1}
\end{equation}
where we have defined a new 
variable,\footnote{
This choice of master variable is important. 
If we use other master variable, 
the $\omega^3,\omega^4, \cdots$ terms may appear in the resultant master
equation.
The good master variable~(\ref{eq:K=1masvar}) can be found by moving
into the Hamiltonian formalism as explained in~\cite{Murata:2008yx}.
}
\begin{equation}
 \Phi_1 = 
\frac{ r^{5/2}( r^4+2\mu a^2)^{5/4}}
{( r^{10}+2\mu a^2 r^6+\mu^2 a^6)^{1/2}}\left[
\left(-i\omega+\frac{4i\mu a}{r^4+2\mu a^2}\right)\frac{h_{r+}}{r^2}
-\left(\frac{h_{t+}}{r^2}\right)'
\right]
\label{eq:K=1masvar}
\end{equation}
and functions $\Omega_{1}$ and $V_1$ are given by
\begin{equation}
 \Omega_1( r) = \frac{2\mu a}{ r^4+2\mu a^2}
\left(1-\frac{a^2  r^4 (5 r^4+6\mu a^2)G}{4( r^{10}+2\mu a^2  r^6+ \mu^2a^6)} \right)
\ ,
\end{equation}
and
\begin{equation}
\begin{split}
V_1( r)=&\,
\frac{G(r)}{4 r^2(2\mu a^2+ r^4)^3( r^{10}+2\mu a^2 r^6+\mu^2a^6)^2}\\
&\times[
15 r^{34}/L^2
+35 r^{32}
+18\mu(1+7a^2/L^2) r^{30}
+310\mu a^2  r^{28}\\
&+8a^2 \mu^2(20+57a^2/L^2) r^{26}
+2 \mu^2a^4(596-75 a^2/L^2) r^{24}\\
&+2 \mu^2 a^4(152\mu+456\mu a^2/L^2-75a^2) r^{22}
+4\mu^3a^6(767-240a^2/L^2) r^{20}\\
&-16\mu^3a^6(8\mu-63\mu a^2/L^2+60a^2) r^{18}
+24\mu^4a^8(217-94 a^2/L^2) r^{16}\\
&-\mu^4a^8(-480\mu a^2/L^2-35a^4/L^2+480\mu+2128a^2) r^{14}\\
&+3\mu^4a^{10}(1424\mu-768\mu a^2/L^2+5a^2) r^{12}
-2\mu^5a^{12}(827-77a^2/L^2) r^{10}\\
&+2\mu^5a^{12}(432\mu-432\mu a^{2}/L^2+25a^2) r^8
-12\mu^6a^{14}(14-15a^2/L^2) r^6\\
&+68\mu^6 a^{16}  r^4
-24\mu^7a^{16}(1-a^2/L^2) r^2
+56\mu^7a^{18}
]\ .
\end{split}
\end{equation}
We have used the tortoise coordinate defined in
Eq.~(\ref{eq:kame}). 

We can get the asymptotic form of $\Omega_1( r)$ and $V_1( r)$ as
\begin{equation}
\Omega_1( r) \rightarrow 0 \quad( r\rightarrow \infty)\ ,\quad
\Omega_1( r) \rightarrow \Omega_H \quad( r\rightarrow  r_+)\ ,
\end{equation}
and
\begin{equation}
 V_{1}( r)\rightarrow 0 \quad( r\rightarrow  r_+)\ ,\quad 
 V_{1}( r)\rightarrow \frac{15r^2}{4L^4} \quad( r\rightarrow \infty)\ ,
\end{equation}
Therefore, asymptotic form of
solution of master equation~(\ref{eq:master_K=1}) becomes
\begin{equation}
\Phi_{1} \rightarrow e^{\pm i\{\omega-2\Omega_H\} r_\ast}
 \quad( r\rightarrow  r_+)\ ,\quad
\Phi_{1} \rightarrow 
r^{-1/2\pm 2}
\quad( r\rightarrow \infty)\ .
\end{equation}

\begin{figure}
\begin{center}
\includegraphics[height=5.5cm,clip]{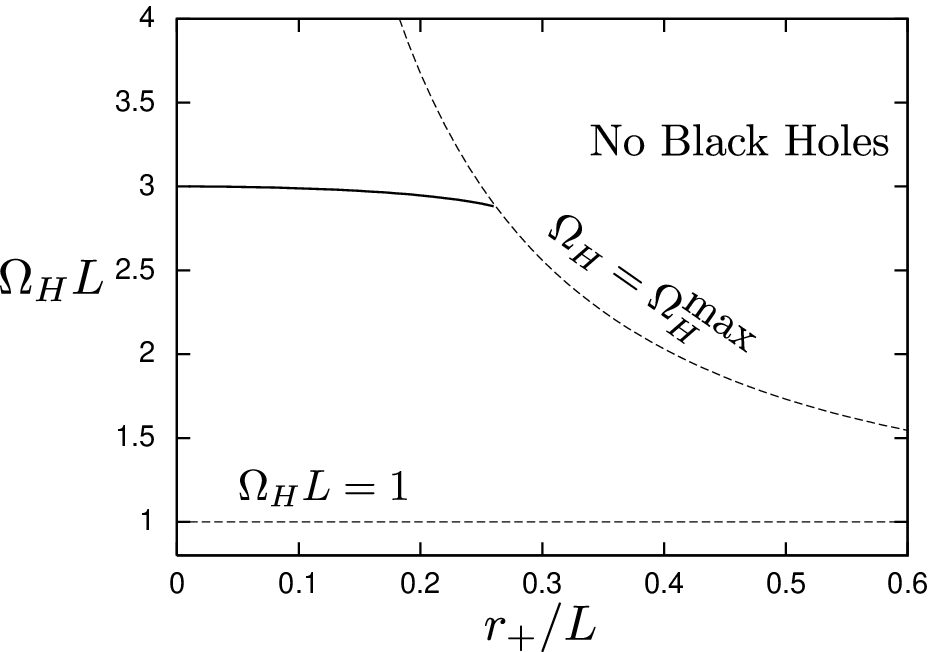}
\caption{\label{fig:SRK=1}The onset line of the superradiant instability 
for $(J=0,M=0,K=1)$ mode is depicted. The solid line is onset of the 
 instability. 
In the region of upper right, there is no black hole
 solutions. 
}
\end{center}
\end{figure}

We can study the onset of superradiant instability of this mode by the
same way as section \ref{sec:KAK>0}. The result is depicted in
Figure.\ref{fig:SRK=1}. We see that onset of the instability is 
$\Omega_HL\simeq 3$. On the other hand, in section \ref{sec:KAK>0}, 
it was shown that the onset is $\Omega_HL=1$ for $(J,M,K=J+2)$ modes. 
Thus, the $(J=0,M=0,K=1)$ mode is irrelevant
for the onset of the instability.

\subsection{$(J,M,K=J+2)$ with $\ell>0$
  modes}\label{app:K=J+2_massive}
\begin{figure}
\begin{center}
\includegraphics[height=5.5cm,clip]{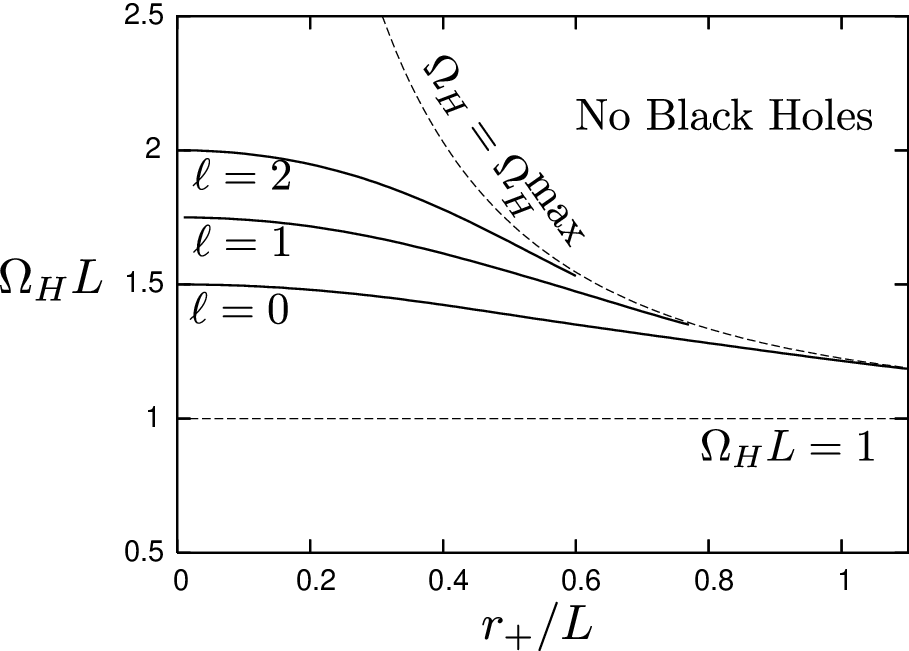}
\caption{\label{fig:SRK=2_massive}The onset line of the superradiant instability 
for $(J=0,M=0,K=2)$ and $\ell=0,1,2$ modes are depicted. 
The solid line is the onset of instability of each modes. 
In the region of upper right, there is no black hole
 solution. 
}
\end{center}
\end{figure}

To see the effect of Kaluza-Klein modes for superradiant instability,
we study the stability of $(J,M,K=J+2)$ with $\ell>0$ modes.
The metric perturbation for these mode is given by
\begin{equation}
 h_{MN}(x^M) dx^M dx^N = h_{++}(r)e^{-i\omega t}
D_{J}(x^i)Y_\ell(\Omega_5)\sigma^+\sigma^+ \ ,
\label{eq:h_K=J+2_massive}
\end{equation}
where $D_J(x^i)\equiv D^J_{K=J,M}(x^i)$ and $Y_\ell(\Omega_5)$ is spherical
harmonics on $S^5$ defined by~(\ref{eq:Y_ell_def}).
We define a new variable as
\begin{equation}
 \Phi = \frac{(r^4+2\mu a^2)^{1/4}}{ r^{3/2}}h_{++}\ .
\end{equation}
Then, using (\ref{eq:masssive}), we can obtain equation for these modes as
\begin{equation}
 -\frac{d^2 \Phi}{d r_\ast^2} + V( r)\Phi =
  [\omega-2(J+2)\Omega( r)]^2 \Phi\ ,
\label{eq:master_K=J+2_massive}
\end{equation}
where functions $\Omega(r)$ and $V(r)$ are determined by
\begin{equation}
 \Omega(r)=\frac{2\mu a}{ r^4 + 2\mu a^2} ,
\end{equation}
and
\begin{equation}
\begin{split}
 V(r)=&\frac{G(r)}{4r^2(r^4+2\mu a^2)^3}\big[
(15+8\varepsilon)r^{14}/L^2+(4J+5)(4J+7)r^{12}\\
&+2\mu(9+33 a^2/L^2+16\varepsilon  a^2/L^2)r^{10}
+2(16 J^2+32J+5)\mu a^2r^8\\
&+(-40+32\varepsilon a^2/L^2+68a^2/L^2)\mu^2a^2r^6
-4\mu^2a^4(16J+35)r^4\\
&+8(1-a^2/L^2)\mu^3a^4r^2
-40\mu^3a^6
\big]\ .
\end{split}
\end{equation}
The equations~(\ref{eq:master_K=J+2_massive}) return to
(\ref{eq:master_K=J+2}) for $\ell=0$.
By the similar way as section~\ref{sebsec:SmallAdS}, 
we can see that, for small black holes, the onset of superradiant
instability is given by $\Omega_HL=(J+3+\ell/2)/(J+2)$.
For any value of $r_+$, we solve (\ref{eq:master_K=J+2_massive})
numerically by the same way as section~\ref{sec:SRnumerical} 
and obtain Figure.\ref{fig:SRK=2_massive}.
We plot the onset of instability for $(J=0,M=0,K=2)$ and $\ell=0,1,2$
modes. From this result, we can see that the superradiant instability of
Kaluza-Klein modes appear at higher $\Omega_H$ than zero mode.
Thus, this result suggest that Kaluza-Klein modes are not relevant to
see the onset of superradiant instability.

\section{Transverse Traceless Condition for Kaluza-Klein Graviton}
\label{app:tt}
In this appendix,
we prove that 
we can impose the transverse traceless condition~(\ref{eq:TT})
in the equation~(\ref{eq:masssive}). 

From the Bianchi identity, we can get
\begin{equation}
\begin{split}
 0&=\delta(g^{\rho\sigma}\nabla_\rho G_{\sigma\mu})\\
&=-h^{\rho\sigma}\nabla_{\rho}G_{\sigma\mu}
+g^{\rho\sigma}(-\delta\Gamma^\lambda_{\rho\sigma}G_{\lambda\mu}
-\delta\Gamma^\lambda_{\rho\mu}G_{\lambda\sigma})
+g^{\rho\sigma}\nabla_\rho\delta G_{\sigma\mu}\ .
\end{split}
\label{eq:delBianchi}
\end{equation}
where $\delta\Gamma^{\rho}_{\mu\nu}$ is perturbation of the Christoffel
symbol defined by
\begin{equation}
 \delta\Gamma^{\rho}_{\mu\nu}
=\frac{1}{2}g^{\rho\sigma}(\nabla_{\mu}h_{\nu\sigma}
+\nabla_\nu h_{\sigma\mu}-\nabla_\sigma h_{\mu\nu})\ .
\label{eq:delGam}
\end{equation}
The back ground equation is given by $G_{\mu\nu}=6L^{-2}g_{\mu\nu}$ and,
thus, we can get 
$\nabla_{\rho}G_{\sigma\mu}=0$.
Therefore, the first
term in the second line of (\ref{eq:delBianchi}) vanishes and we can get 
\begin{equation}
g^{\rho\sigma}\nabla_\rho\delta G_{\sigma\mu}=
6L^{-2}(g^{\rho\sigma}g_{\lambda\mu}\delta\Gamma^\lambda_{\rho\sigma}
+\delta\Gamma^\rho_{\rho\mu})
=6L^{-2}\nabla^\rho h_{\rho\mu}\  ,
\end{equation}
where we have used the expression~(\ref{eq:delGam}) at the last equality.
Hence, from the divergence of the perturbation equation~(\ref{eq:masssive}), 
\begin{equation}
\nabla^\rho(h_{\rho\mu}-\frac{1}{2}g_{\rho\mu}h)
=-\varepsilon^{-1}L^2\nabla^\rho(\delta G_{\rho\mu}-6L^{-2}h_{\mu\nu})
=0\ .
\label{eq:constraint}
\end{equation}
It is the constraint equation of~(\ref{eq:masssive}).

Now, we consider the trace of~(\ref{eq:masssive}). The trace of
perturbation of the Einstein tensor is given by
\begin{equation}
 g^{\rho\sigma}\delta G_{\rho\sigma}=
-\frac{3}{2}\nabla^\rho\nabla^\sigma(h_{\rho\sigma}-g_{\rho\sigma}h)
+\frac{5}{2}h_{\rho\sigma}(R^{\rho\sigma}-\frac{1}{5}g^{\rho\sigma}R)\ .
\label{eq:trG}
\end{equation}
Because of $R_{\mu\nu}=-4L^{-2}g_{\mu\nu}$ and $R=-20L^{-2}$, 
Ricci tensor and Ricci scalar terms in~(\ref{eq:trG}) are cancelled each
other. Thus, making use of the constraint equation~(\ref{eq:constraint}), 
the equation (\ref{eq:trG}) becomes
\begin{equation}
 g^{\rho\sigma}\delta G_{\rho\sigma}=\frac{3}{4}\nabla^2 h\ .
\end{equation}
Thus, from the trace of~(\ref{eq:masssive}), we can get the equation for
trace part of $h_{\mu\nu}$ as
\begin{equation}
 \nabla^2 h = 2(\varepsilon+2)L^2 h\ .
\end{equation}
Therefore, trace part of $h_{\mu\nu}$ is decoupled from other components of
$h_{\mu\nu}$ and we can put $h=0$ consistently. 
Then, constraint equation becomes $\nabla^\rho h_{\rho\mu}=0$.


\end{document}